\documentclass[a4paper,fleqn,usenatbib]{mnras}
\usepackage[T1]{fontenc}
\usepackage{ae,aecompl}
\usepackage{graphicx}
\usepackage{amsmath}
\usepackage{amssymb}
\usepackage{txfonts}
\usepackage{bbold}
\usepackage{upgreek}

\usepackage{float}
\usepackage{adjustbox}

\graphicspath{ {images/} }

\newcommand{\ltsima}{$\; \buildrel < \over \sim \;$}
\newcommand{\lsim}{\lower.5ex\hbox{\ltsima}}
\newcommand{\gtsima}{$\; \buildrel > \over \sim \;$}
\newcommand{\gsim}{\lower.5ex\hbox{\gtsima}}

\newcommand{\dd}{\mathrm{d}}

\onecolumn

\title[Intrinsic and extrinsic flexions]
{Intrinsic and extrinsic gravitational flexions}
\author[E.S. Giesel, B. Ghosh, B.M. Sch{\"a}fer]
{Eileen Sophie Giesel$^1$\thanks{e-mail: e.giesel@stud.uni-heidelberg.de}, Basundhara Ghosh$^2$\thanks{e-mail: basundharag@iisc.ac.in}, Bj{\"o}rn Malte Sch{\"a}fer$^1$\thanks{e-mail: bjoern.malte.schaefer@uni-heidelberg.de}\\
$^1$Zentrum f{\"u}r Astronomie der Universit{\"a}t Heidelberg, Astronomisches Rechen-Institut, Philosophenweg 12, 69120 Heidelberg, Germany\\
$^2$Department of Physics, Indian Institute of Science, C. V. Raman Road, Bangalore 560012, India
}

\begin{document}
\pagerange{\pageref{firstpage}--\pageref{lastpage}}
\pubyear{2021}
\maketitle
\label{firstpage}

\begin{abstract}
The topic of this paper is a generalisation of the linear model for intrinsic alignments of galaxies to intrinsic flexions: In this model, third moments of the brightness distribution reflect distortions of elliptical galaxies caused by third derivatives of the gravitational potential, or, equivalently, gradients of the tidal gravitational fields. With this extension of the linear model mediating between the brightness distribution and the tidal gravitational fields and with a quantification of the shape of the galaxy at third order provided by the HOLICs-formalism, we are able to compute angular spectra of intrinsic flexions and the cross-spectra with weak lensing flexions. Spectra for intrinsic flexions are typically an order of magnitude smaller than lensing flexions, exactly as in the case of intrinsic ellipticity in comparison to weak shear. We find a negative cross correlation between intrinsic and extrinsic gravitational flexions, too, complementing the analogous correlation between intrinsic and extrinsic ellipticity. After discussing the physical details of the alignment model to provide intrinsic flexions and their scaling properties, we quantify the observability of the intrinsic and extrinsic flexions and estimate with the Fisher-formalism how well the alignment parameter can be determined from a Euclid-like weak lensing survey. Intrinsic flexions are very difficult to measure and yield appreciable signals only with highly optimistic parameter choices and noise levels, while being basically undetectable for more realistisc flexion measurement errors.
\end{abstract}

\begin{keywords}
gravitational lensing: weak -- dark energy -- large-scale structure of Universe.
\end{keywords}

\section{introduction}\label{sect_intro}
Intrinsic alignments refer to shape correlations of galaxies, likely caused by tidal interaction with the cosmic large scale structure \citep[see][for reviews]{joachimi_galaxy_2015, kirk_galaxy_2015,kiessling_galaxy_2015, troxel_intrinsic_2015}. They are considered to be an important systematic in weak gravitational lensing, as they contaminate angular spectra of weak shear to a level amounting to about 10\% over a wide range of multipoles \citep{hirata_intrinsic_2010}. Different models for the detailed mechanism that brings about intrinsic alignments are being discussed in the literature and commonly differentiate between different galaxy types: While the alignment of spiral galaxies could be related to the orientation of the angular momentum, which in turn is built-up in by tidal torquing in the early stages of galaxy formation \citep{natarajan_angular_2001, crittenden_spin-induced_2001, lee_nonlinear_2007, schafer_galactic_2012, capranico_intrinsic_2013, codis_spin_2015, schaefer_angular_2015}, the alignment of elliptical galaxies is thought to be generated by direct tidal interaction of the galaxy with gravitational fields sourced by the surrounding large-scale structure \citep{hirata_galaxy-galaxy_2004, hirata_intrinsic_2010}, where they ultimately arise as terms in a perturbative expansion \citep{blazek_beyond_2017, fang_fast-pt_2017, vlah_eft_2019}.

Using this model, it has been possible to predict correlations of the shapes of galaxies with analytic formalisms \cite{blazek_beyond_2017, blazek_separating_2012, schmitz_time_2018}, and there is evidence for this mechanism in numerical simulations of galaxy formation \cite{tenneti_galaxy_2014, codis_intrinsic_2015, hilbert_intrinsic_2016,Zjupa_tng_2020,chisari_intrinsic_2015,chisari_redshift_2016, codis_galaxy_2018} as well as positive evidence from actual surveys \cite{kilbinger_cfhtlens:_2013, heymans_cfhtlens_2013, jee_cosmic_2015, joudaki_cfhtlens_2017, johnston_kids+gama:_2018}.

Weak lensing flexions denote the generation of third moments of the brightness distribution of galaxies in weak gravitational lensing \citep{bacon_weak_2006, goldberg_measuring_2007, schneider_weak_2008}, if there are appreciable third derivatives of the lensing potential, or, equivalently, gradients of lensing shear and convergence across the galaxy image \citep{fluke_shape_2011}. As such, they echo at higher order the idea that ellipticities, which reflect the second moments of the brightness distribution, are changed by lensing shear and convergence. A straightforward way to quantify distortions of galaxies at third order are for instance HOLICs proposed by \citet{okura_new_2007, okura_method_2008}, which are sensible estimates of the third derivatives of the lensing potential, or rather the modes of that quantity. While we prefer to work with HOLICs in this paper, there are of course alternative quantification of the flexion distortion \citep{massey_weak_2007}.

Lensing flexions are a weak effect and correspondingly, there are only detection reports at comparatively modest levels of statistical significance, which is in strong contrast to weak shear. Applications of lensing flexions are not straightforward to identify: In a statistical quantification analogous to weak cosmic shear one needs to emphasise that weak flexion spectra carry exactly the same information as weak shear spectra \citep{bacon_weak_2006}, albeit come with a higher amount of shape noise. 

Compared to lensing shear and convergence, flexions are naturally a small-scale effect as they arise as third derivatives instead of second derivatives of the lensing potential, or, in the case of intrinsic alignments, of the actual physical potential at the location of the galaxy. This property might be advantageous if small-scale information of the cosmic density field contains valuable information, for instance on non-Gaussianity \citep{munshi_higher_2011, munshi_higher_2011-1, fedeli_constraining_2012}, on higher-order lensing corrections \citep{schafer_validity_2012}, on halo shapes \citep{lasky_shape_2009, hawken_gravitational_2009, er_estimate_2010} and halo reconstructions \citep{er_mass_2010}, on the amount of substructure in dark matter haloes \citep{velander_probing_2010, rowe_new_2007, leonard_gravitational_2007, bacon_measuring_2010}, for de-lensing applications \citep{shapiro_delensing_2010}, or in galaxy-galaxy lensing \citep{goldberg_galaxy-galaxy_2005}. Whether higher-order effects of gravity manifest themselves in lensing on the flexion-level remains unclear \citep{sereno_aberration_2008}.

The question which motivated our paper concerns the existence and the physical properties of intrinsic flexions, i.e. of a shape distortion of a galaxy of the flexion type which is caused by direct tidal interaction of the galaxy with its environment, specifically with the third derivatives of the gravitational potentials. For this purpose, we construct a model for tidal interaction of elliptical galaxies, where the third moments of the perturbed brightness distribution reflects the magnitude of the third derivatives of the gravitational potential, or equivalently, the gradients of the tidal shear. While the information content of weak shear and weak flexion spectra are identical, which applies to likewise the effect evoked by gravitational lensing and by intrinsic tidal interaction alike, we investigate whether intrinsic flexions would be observable by future surveys and how they would depend on the internal structure of galaxies, for instance on the S{\'e}rsic profile family or on the velocity dispersion of the galaxy.

After generalising tidal interactions of elliptical galaxies with their surrounding large-scale structure to third order in Sect.~\ref{sect_tidal}, we illustrate how the resulting brightness distributions from direct tidal interaction can be quantified in analogy to those from gravitational lensing by using the HOLICs-formalism \citep{okura_new_2007, okura_method_2008}, resulting in angular spectra for the intrinsic and extrinsic flexion distortion shown in Sect.~\ref{sect_spectra}. We quantify the observability of intrinsic and extrinsic flexion correlations in a Euclid-like weak lensing survey \citep{Amendola:2016saw} in Sect.~\ref{sect_s2n} before moving on to cosmological constraints in Sect.~\ref{sect_fisher}. We summarise our results and discuss limitations to our approach in Sect.~\ref{sect_summary}. Throughout the paper we work in the context of a $w$CDM-cosmology with a constant equation of state value of $w$ close to $-1$ ($w=-0.9$), and standard values for the cosmological parameters, i.e. $\Omega_m = 0.3$, $\sigma_8 =  0.8$, $h = 0.7$ and $n_s = 0.96$, and a parameterised spectrum for nonlinearly evolving scales. Throughout the paper, summation convention is implied.

\section{tidal interactions of galaxies and gravitational lensing}\label{sect_tidal}
Linear tidal interaction models have first been introduced by \citet{hirata_galaxy-galaxy_2004, hirata_intrinsic_2010}. They are supposed to explain a linear relationship between the ellipticity of a galaxy and the tidal gravitational field through direct tidal deformation. The physical relationship in the parametrisation of linear models has been clarified by \citet{piras_mass_2018, ghosh2020intrinsic}, who discuss the scaling of the alignment with galaxy properties, most importantly the velocity dispersion as a reflection of the depth of the gravitational potential. In this work we extend this particular model to find the according intrinsic flexions, and to correlate them with the lensing flexions. Hence, the derivation will mainly follow the arguments made by \citet{piras_mass_2018, ghosh2020intrinsic}: The main idea is that galaxies should experience intrinsic shape deformations for they are subject to tidal gravitational fields sourced by the ambient matter distribution, leading to intrinsic shape and size fluctuations. However, the tidal fields themselves also experience variations, which are manifested in third order variations of the stellar density, and hence higher order shape distortions, specifically shape and size gradients. Ultimately, one would expect on the basis of the linear alignment model, perturbations of the $n$th moments of the brightness distribution as a consequence of non-vanishing $n$th derivatives of the gravitational potential.

To this purpose, we model the tidal effect on a virialised, self-gravitating and spherical system with constant velocity dispersion $\sigma^2$ and vanishing averaged velocity in Jeans-equilibrium. Then, the relation between stellar density and gravitational potential as a function of distance $r$ from the minimum of the potential is given by
\begin{equation}
\sigma^2 \frac{\dd}{\dd r}\ln\rho = - \frac{\dd}{\dd r}\Phi\,.
\end{equation}
Consequently, the stellar density of the unperturbed system is given by
\begin{equation}\label{eq:JeansDensity}
\rho(r) \varpropto  \exp\left( - \frac{\Phi(r)}{\sigma^2}  \right)\,,
\end{equation}
such that the density $\rho$ assumes the highest value at the center of the galaxy, where the potential is minimized. Now, the stellar density for perturbations in the potential can be derived up to third order, similar to \citet{ghosh2020intrinsic}: Under the gravitational influence of surrounding galaxies one can expand the gravitational field to third order at the origin to receive
\begin{equation}\label{eq:perturbedPotential}
\Phi(r) \, \rightarrow \,  
\Phi(r) +
\frac{1}{2!} \partial_{a} \partial_{b} \Phi r^a r^b + 
\frac{1}{3!} \partial_{a} \partial_{b} \partial_c \Phi r^a r^b r^c\,.
\end{equation}
Now, inserting the expansion~(\ref{eq:perturbedPotential}) into equation~(\ref{eq:JeansDensity}), assuming that the tidal field and its derivative are small, a first-order Taylor expansion with respect to the higher order derivatives of $\Phi$ leads to the following perturbed density profile:
\begin{equation}\label{eq:Density}
\rho(r) \rightarrow 
\exp\left( - \frac{\Phi(r)}{\sigma^2}  \right) 
\left(1- \frac{1}{2! \sigma^2} \partial_{a} \partial_{b} \Phi r^a r^b - \frac{1}{3! \sigma^2} \partial_{a} \partial_{b} \partial_c \Phi r^a r^b r^c \right) \equiv 
\rho(r) \left(1- \frac{1}{2! \sigma^2} \Phi_{ab} r^a r^b - \frac{1}{3! \sigma^2} \Phi_{abc} r^a r^b r^c \right)\,,
\end{equation}
where the notations $\partial_{a} \partial_{b} \Phi \equiv \Phi_{ab}$ and $\partial_{a} \partial_{b} \partial_c \Phi \equiv \Phi_{abc}$ are introduced for simplicity.

To quantify the effect of $\Phi_{abc}$ onto the intrinsic shape of the galaxy one needs to evaluate the change in the octupole moment of the surface brightness distribution $q_{abc} \varpropto  \int \mathrm{d}^2 r \rho(r) r_a r_b r_c\,$, similar to the work by \citet{ghosh2020intrinsic}, where the tidal field $\Phi_{ab}$ led to perturbations in the quadrupole moment $q_{ab} \varpropto  \int \mathrm{d}^2 r \rho(r) r_a r_b \,$ of the surface brightness distribution.

As done by \citet{ghosh2020intrinsic} one also assumes that the surface brightness distribution is proportional to the stellar density of the galaxy, hence approximating the higher order moments of the surface brightness distribution with higher order moments of a $2$-dimensional projected stellar density distribution. Hereby, it is neglected that one would actually need to derive the projected stellar density from the unprojected $3$-dimensional density via an Abel transform (for Ref. see \citet{Arfken05}), which would lead to a numerical correction prefactor dropping out due to normalisation. Thus, using polar coordinates $r_0 = r \cos(\phi)$ and $r_1= r \sin(\phi)$ such that the indices run from $0$ to $1$, the change in the octupole moment becomes
\begin{equation}\label{eq:thirdmoment}
\begin{split}
\Delta q_{abc} =&\int \mathrm{d}^2 r \rho(r) r_a r_b r_c \left(  r_d r_e \frac{\Phi_{de}}{2 \sigma^2} +  r_d r_e r_f \frac{\Phi_{def}}{6 \sigma^2} \right)
\
\\
=& \frac{1}{2 \sigma^2}  \int_{0}^{\infty} \mathrm{d} r r^6 \rho(r) \int_{0}^{2 \uppi}  \mathrm{d} \phi \cos^{5-(a+b+c+d+e)}\phi \sin^{a+b+c+d+e} \phi\,\Phi_{de}
+ \frac{1}{6 \sigma^2}  \int_{0}^{\infty} \mathrm{d} r r^7 \rho(r)\int_{0}^{2 \uppi}  \mathrm{d} \phi \cos^{6-(a+b+c+d+e+f)}\phi \sin^{a+b+c+d+e+f} \phi \,\Phi_{def}
\
\\
=& D_{\mathcal{F}} S_{abcdef} \Phi_{def} \,,
\end{split}
\end{equation}
where we introduced the notation $S_{abcde} \equiv \int_{0}^{2 \uppi}  \mathrm{d} \phi \cos^{5-(a+b+c+d+e)}\phi \sin^{a+b+c+d+e} \phi  = 0$ and $S_{abcdef} \equiv \int_{0}^{2 \uppi}  \mathrm{d} \phi \cos^{6-(a+b+c+d+e+f)}\phi \sin^{a+b+c+d+e+f} \phi$ for the angular integrals, while the purely radial integral is denoted by $D_{\mathcal{F}} \equiv 1/(6 \sigma^2) \int_{0}^{\infty} \mathrm{d} r r^7 \rho(r)$. 
The value of the angular integral $S_{abcde}$ vanishes for any combination of indices, which shows that the octupole moment variations are independent of the tidal field $\Phi_{ab}$, in the same way as the quadrupole moments would remain uninfluenced by the third derivatives $\Phi_{abc}$: As such the distortions modes are mutually independent, as a consequence of the linearity of the model. Non-vanishing values for $S_{abcdef}$ are the contributions $S_{000000}=S_{111111}= 5\uppi/8$ and $S_{000011}=S_{111100} = \uppi/8$ with any index permutation yielding the same number, due to the commutativity of the index sum.

The prefactor $D_{\mathcal{F}} S_{abcdef} \varpropto D_{IA,\mathcal{F}}$ of eqn.~(\ref{eq:thirdmoment}) measures the sensitivity of the reaction of the shape to variations in the tidal field, containing information about the alignment parameter for flexions denoted by $D_{IA,\mathcal{F}}$, similar to the alignment parameter $D_{IA}$ discussed in \citet{ghosh2020intrinsic} for intrinsic sizes and ellipticities. However one needs to take one subtlety into account concering the numerical value of the alignment parameter: As already shown by \cite{ghosh2020intrinsic} the alignment parameter scales with $r_{\text{scale}}^2$ as scale for the galaxy radius squared. For elliptical galaxies with a Sérsic profile this length scale corresponds to the Sérsic radius $r_S \approx 1 \text{kpc}$. However, in this work we are considering virialized systems by imposing the Jeans-equilibrium, and thus we would need to rescale the alignment parameter with the virial radius $r_{\text{vir}}\approx 100 \text{kpc}$ squared as measure of the galaxy size in this model. Thus we need to rescale $D_{IA}$ by a factor $r_{\text{vir}}^2/r_S^2\approx 10^4$. Thus, a lower limit for the numerical value of the alignment parameter in our analysis is given by $D_{IA} \varpropto - 10^{-2} \text{Mpc}/\text{h}$ with which we proceed in the following.

By decomposing the rank-3 object $\Phi_{abc}$ into an orthonormal basis of $4\times4$-matrices $\Delta^{(n)}$ in a block diagonal matrix representation $\left(\Phi\right)_{b+2a, c+2a} =\Phi_{abc}$, one can extract the linearly independent contributions to the intrinsic flexion. This is similar to previous works by \citet{ghosh2020intrinsic} where the tidal field was decomposed using the Pauli matrices $\sigma^{(n)}$ to separate the effects of tidal shear on size and ellipticity. For intrinsic flexions the appropriate basis decomposition is given by a similar set of matrices, derived from the Dirac matrices
\begin{align}\label{eq:Diracs}
\text{spin-1 type:} \,\, \Delta^{(1)} = \frac{1}{\sqrt{3}} \begin{pmatrix} 2 \sigma^{(0)} + \sigma^{(3)} & 0 \\ 0 & \sigma^{(1)}    \end{pmatrix}, \, \Delta^{(2)} = \frac{1}{\sqrt{3}} \begin{pmatrix} \sigma^{(1)} & 0 \\ 0 & 2 \sigma^{(0)} - \sigma^{(3)}\end{pmatrix},
\quad
\text{spin-3 type:} \,\, \Delta^{(3)} = \begin{pmatrix} \sigma^{(3)} & 0 \\ 0 & -\sigma^{(1)}    \end{pmatrix}, \,  \Delta^{(4)}= \begin{pmatrix} \sigma^{(1)} & 0 \\ 0 & \sigma^{(3)}    \end{pmatrix}.
\end{align}
as shown in Appendix~\ref{appA} from the decomposition of the lensing flexion. This idea is similar to \citet{schafer_validity_2012}, however, here the tensor $\Phi_{abc}$ is symmetric in all its three indices, such that it has four independent degrees of freedom, unlike in \citet{schafer_validity_2012} where dropping Born's approximation leads to six independent components of the lensing flexion instead of four. The Pauli matrices $\sigma^{(n)}$ are given by
\begin{equation}
\sigma^{(0)} = \begin{pmatrix} \,1 & \,0 \\ \,0&\,1 \end{pmatrix}, \, \sigma^{(1)} = \begin{pmatrix} \,0&\,1 \\ \,1&\,0 \end{pmatrix}, \sigma^{(2)} = \begin{pmatrix} \,0 & -\mathrm{i} \\ \,\mathrm{i}&\,0 \end{pmatrix},  \, \sigma^{(3)} = \begin{pmatrix} \,1 &\,0 \\ \,0&-1 \end{pmatrix}.
\end{equation}
The spin-1 type distortions $\zeta_1$ and $\zeta_2$ of the galaxy shape proportional to $\Phi_{abc} \varpropto \Delta^{(1)}$ and $\Phi_{abc} \varpropto \Delta^{(2)}$ measure an intrinsic centroid shift and can be estimated by
\begin{equation}\label{eq:zeta1}
\zeta_1 =  \frac{1}{4}\Delta q_{abc}\Delta^{(1)}_{abc} \varpropto \frac{1}{4}S_{abcdef}\Delta^{(1)}_{def}\Delta^{(1)}_{abc} =\frac{3 \uppi}{4}\,, \quad \text{and} \quad \zeta_2 =  \frac{1}{4}\Delta q_{abc}\Delta^{(2)}_{abc}\varpropto\frac{1}{4}S_{abcdef}\Delta^{(2)}_{def}\Delta^{(2)}_{abc} =\frac{3 \uppi}{4}\,.
\end{equation}
Physically, they can be interpreted as a gradient in convergence or in intrinsic size. Similarly, the spin-3 type distortions establishing a three-fold symmetry in the galaxy shape, $\delta_1$ and $\delta_2$ are given by
\begin{equation}\label{eq:delta1}
\delta_1 = \frac{1}{4} \Delta q_{abc}\Delta^{(3)}_{abc} \varpropto \frac{1}{4}S_{abcdef}\Delta^{(3)}_{def}\Delta^{(3)}_{abc} =\frac{\uppi}{4}\,, \quad \text{and} \quad \delta_2 = \frac{1}{4} \Delta q_{abc}\Delta^{(4)}_{abc} \varpropto \frac{1}{4}S_{abcdef}\Delta^{(4)}_{def}\Delta^{(4)}_{abc} =\frac{\uppi}{4}\,.
\end{equation}
Again, a physical interpretation would be a gradient in the shear field, or in the intrinsically induced ellipticity. 

We conclude that the changes $\delta_n$ in the observables due to spin-3 type distortions are one third of the ones in $\zeta_n$ caused by the parts of $\Phi_{abc}$ which are proportional to the spin-1 type fields. Now, to quantify the respective relative change of the galaxy shape compared to the unperturbed shape one can use the so-called HOLICs (see \citet{okura_new_2007, okura_method_2008} for reference) quantifying the normalised changes of the octupole moment $\Delta q_{abc}$ of the surface brightness distribution $I\left(\theta\right)\varpropto \rho(r)$.

According to  \citet{okura_new_2007} the $\zeta$-HOLIC with spin-1 and the $\delta$-HOLIC with spin-3 symmetry can be obtained by
\begin{equation}
\zeta = 
\frac{\left(q_{000} + q_{011} \right) + \mathrm{i} \left(q_{001} + q_{111}  \right)}{\xi}\,, 
\quad \text{and} \quad 
\delta = 
\frac{\left(q_{000} - 3 q_{011} \right) + \mathrm{i} \left(3 q_{001} - q_{111}  \right)}{\xi}\,.
\end{equation}
The normalisation factor $\xi$ is the spin-0 component of the hexadecupole $q_{abcd} \varpropto \int \mathrm{d}^2 r \rho\left(r\right) r_a r_b r_c r_d$ of the unperturbed density distribution, which is  $\xi = q_{0000} +2 q_{0011} + q_{1111}$ (again, due to index exchange symmetry) such that $\zeta$ and $\delta$ are given in units of inverse length and inverse angle respectively, identically to the lensing flexions.

With these definitions one can now derive the change of the intrinsic HOLICs under the influence of variations of the tidal fields, and relate these to the decomposition of $\Phi_{abc}$ in terms of the matrices given in (\ref{eq:Diracs}) and the intrinsic distortion quantities $\zeta_1$, $\zeta_2$ (\ref{eq:zeta1}) and $\delta_1$, $\delta_2$ (\ref{eq:delta1}).

Therefore, the intrinsic flexion with spin-1, expressed by the change in the $\zeta$-HOLIC due to variations in the tidal field is given by
\begin{equation}\label{eq:spin1holic}
\begin{split}
\Delta \zeta = 
\frac{\left(\Delta q_{000}+  \Delta q_{011}\right)+ \mathrm{i}\left(\Delta q_{001}+ \Delta q_{111}\right)}{\xi} = \frac{1}{\sqrt{3}} \frac{\left(\Delta q_{abc} \Delta^{(1)}_{abc}\right) + \mathrm{i}\left(\Delta q_{abc} \Delta^{(2)}_{abc}\right) }{\xi} = \frac{4}{\sqrt{3}} \frac{ \zeta_1 + \mathrm{i}\zeta_2 }{\xi}\,,
\end{split}
\end{equation}
while for the intrinsic spin-3 flexion one obtains
\begin{equation}\label{eq:spin3holic}
\begin{split}
\Delta \delta = \frac{\left(\Delta q_{000} - 3\Delta q_{011}\right)+ \mathrm{i}\left( 3\Delta q_{001}-  \Delta q_{111}   \right)}{\xi}= \frac{\left(\Delta q_{abc} \Delta^{(3)}_{abc}\right) + \mathrm{i} \left(\Delta q_{abc} \Delta^{(4)}_{abc}\right) }{\xi} =4\frac{\delta_1 + \mathrm{i} \delta_2 }{\xi}\,.
\end{split}
\end{equation}

Using the relations~(\ref{eq:spin1holic}) and~(\ref{eq:spin3holic}) we explicitly estimate the radial part of the alignment parameter $D_{IA,\mathcal{F}}\varpropto D_\mathcal{F} /\xi$ for the flexion as the susceptibility between third order shape deformations and third order perturbations in the potential.

The changes in the octupole moments of the brightness distribution $\Delta q_{abc}$ are proportional to $D_\mathcal{F}\varpropto\int_{0}^\infty \mathrm{d}r \rho(r) r^7 $, while the normalisation factor $\xi$ as the hexadecupole moment of the unperturbed stellar density distribution one receives
\begin{equation}
\begin{split}
\xi &= q_{0000}+2q_{0011}+q_{1111} = \int \mathrm{d}^2r \rho(r) \left(r_0^4+2r_0^2r_1^2+r_1^4\right)  = \int \mathrm{d}^2r \rho(r) r^4 = 2 \uppi \int_{0}^\infty \mathrm{d}r \rho(r)r^5,
\end{split}
\end{equation}
such that the radial part of the flexion susceptibility factor is in total given by 
\begin{equation}
D_{IA,\mathcal{F}}\varpropto \frac{D_\mathcal{F} }{\xi} \varpropto \frac{\int_{0}^\infty \mathrm{d}r \rho(r) r^7}{\int_{0}^\infty \mathrm{d}r \rho(r)r^5} \equiv \tilde{D}_\mathcal{F}.
\end{equation}

Now, using the S{\'e}rsic-Profile (for details see \citet{sersic_influence_1963, graham_concise_2005, de_vaucouleurs_recherches_1948}) to describe the stellar density $\rho(r)$, similarly to the previous work by \citet{ghosh2020intrinsic},
\begin{equation}\label{eq:sersic}
\rho(r) \varpropto \exp\left(-b(n) \left[\left(\frac{r}{r_{\text{scale}}}\right)^{n^{-1}} -1   \right] \right)\, 
\end{equation}
with $b(n) \approx 2n -\frac{1}{3}$, and $n$ as the S{\'e}rsic index, we receive 
\begin{equation}
\tilde{D}_\mathcal{F} =  \frac{\displaystyle{\int_{0}^\infty} \mathrm{d}r \rho(r) r^7}{\displaystyle{\int_{0}^\infty} \mathrm{d}r \rho(r)r^5} =r_{\text{scale}}^2\, \frac{\displaystyle{\int_{-b}^\infty} \mathrm{d}x \left(x/b+1  \right)^{8n-1} \exp(-x)}{\displaystyle{\int_{-b}^\infty} \mathrm{d}x \left(x/b+1  \right)^{6n-1} \exp(-x)}= r_{\text{scale}}^2 \frac{b^{6n-1}}{b^{8n-1}} \frac{\displaystyle{\int_{0}^\infty} \mathrm{d}y \, y^{8n-1} \exp(-y)}{\displaystyle{\int_{0}^\infty} \mathrm{d}y \, y^{6n-1} \exp(-y)}  = r_{\text{scale}}^2 b^{-2n} \frac{\Gamma(8n)}{\Gamma(6n)} \,.
\end{equation}

Here, the two substitutions $x=b\left[\left(r / r_{\text{scale}}\right)^{n^{-1}}-1 \right]$ from \citet{ghosh2020intrinsic} and additionally $y=x+b$ are used to express the integrals analytically in terms of the $\Gamma(n)$-function and $r_{\text{scale}}$ is the scale radius for the typical size of an elliptical galaxy. In a similar manner one can now rewrite the susceptibility factor $\tilde{D}_\epsilon$ for the relative change of the intrinsic ellipticities derived by \citet{ghosh2020intrinsic} as
\begin{equation}
\Delta \epsilon = \frac{\Delta q_{00} - \Delta q_{11}}{q_{00}+q_{11}}+ 2 \mathrm{i}\frac{\Delta q_{01}}{q_{00}+q_{11}} \varpropto \frac{\int_{0}^\infty \mathrm{d}r \rho(r) r^5}{\int_{0}^\infty \mathrm{d}r \rho(r) r^3} = r_{\text{scale}}^2 b^{-2n} \frac{\Gamma(6n)}{\Gamma(4n)} \equiv  \tilde{D}_\epsilon\,, 
\end{equation}
with unperturbed size $s_0 \varpropto q_{00}+q_{11}= \int \mathrm{d}^2 r \rho(r) \left(r_0^2+r_1^2\right) = 2 \uppi \int_{0}^\infty \mathrm{d}r \rho(r) r^3$.
If we now compare the alignment parameter $D_{IA} \varpropto \tilde{D}_\epsilon$ for the intrinsic ellipticities with the alignment parameter $D_{IA,\mathcal{F}} \varpropto \tilde{D}_\mathcal{F}$ for instrinsic flexions we find the susceptibility ratio
\begin{equation}\label{eq:prefactor}
\frac{D_{IA,\mathcal{F}}}{D_{IA}} = \frac{2 \sigma^2}{6 \sigma^2} \frac{\tilde{D}_\mathcal{F} S_{abcdef}}{\tilde{D}_\epsilon S_{abcd}} =\frac{1}{6} \frac{\tilde{D}_\mathcal{F} }{\tilde{D}_\epsilon}\,, \quad \text{with} \quad \frac{\tilde{D}_\mathcal{F} }{\tilde{D}_\epsilon} = \frac{\Gamma(8n)\Gamma(4n)}{\Gamma(6n)^2}.
\end{equation}
which only depends on the S{\'e}rsic-index $n$. Here the value for the angular integral $S_{abcdef} \varpropto \uppi/4$ from relation (\ref{eq:delta1}) was used while \citet{ghosh2020intrinsic} derived $S_{abcd} \varpropto \uppi/2$ for the angular integral for the intrinsic ellipticites. Also $D_{IA,\mathcal{F}}\varpropto 1/6 \sigma^2$ while $D_{IA}\varpropto 1/2 \sigma^2$ due to the expansion of the density profile in equation (\ref{eq:Density}).
Hence, for small S{\'e}rsic indices around $n \leq 4$, which correspond to usual elliptical galaxies, the quotient of the two different alignment parameters is less than of order $10$.
\begin{figure}
\centering
\includegraphics[scale=0.35]{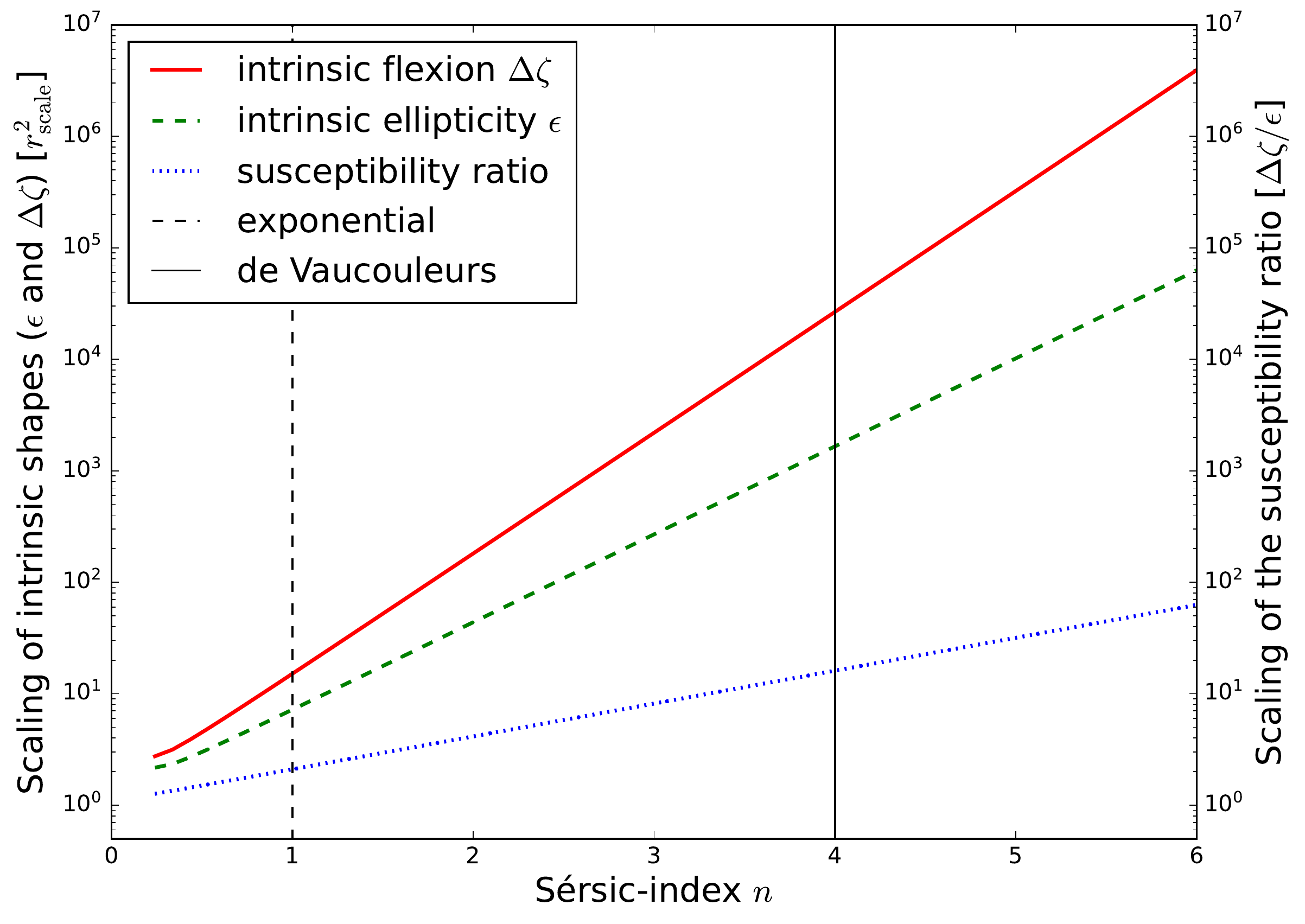}
\caption{Scaling of the relation of the intrinsic flexion $\Delta\zeta$ and of the intrinsic ellipticity $\epsilon$ with S{\'e}rsic-index $n$, for a given tidal gravitational field and a fixed velocity dispersion $\sigma^2$. As particular cases, the exponential profile ($n=1$) and the de Vaucouleurs-profile ($n=4$) are indicated by vertical lines.}
\label{fig_sersic_scaling}
\end{figure}

Fig.~\ref{fig_sersic_scaling} shows the scaling of the response of a galaxy to external tidal fields, specifically to the second and third derivatives of the gravitational potential, giving rise to shape and flexion distortions, respectively. The numerical results suggest that the intrinsic flexion depends more strongly on S{\'e}rsic-index $n$ than intrinsic ellipticity, but that for exponential profiles corresponding to $n = 1$ the values of the effective alignment parameter are similar in magnitude. Additionally, we plot the ratio $\tilde{D}_\mathcal{F} /\tilde{D}_\epsilon$  between the two alignment parameters, which we refer to as the susceptibility ratio, into the same plot, illustrating again the stronger dependence of intrinsic flexion to the S{\'e}rsic-index compared to intrinsic ellipticity.

Interestingly, one can further relate the intrinsic sizes and ellipticities derived in \citet{ghosh2020intrinsic} to intrinsic flexions, by showing that the latter are just variations of the intrinsic size and ellipticities using the alignment model. For the relative size change $\Delta s = s/s_0$ one obtains
\begin{equation}
\Delta s = 
\frac{1}{2 s_0} \Delta q_{cd} \sigma^{(0)}_{cd} =  \frac{1}{2 s_0}  \int \mathrm{d}^2 r \rho(r) r_a r_b \frac{\Phi_{ab}}{2 \sigma^2} r_c r_d \sigma^{(0)}_{cd} = \frac{\int_0^{2 \uppi} \mathrm{d} \phi \int \mathrm{d} r r^5 \rho(r) \left( \cos^2\left(\phi\right) \Phi_{0 0}+ 2 \cos\left(\phi\right)\sin\left(\phi\right) \Phi_{0 1} + \sin^2\left(\phi\right) \Phi_{11}\right)}{4 \sigma^2 \cdot 2 \uppi  \int \mathrm{d} r r^3 \rho(r)} =\frac{1}{8 \sigma^2} \tilde{D}_\epsilon \Delta \Phi\,.
\end{equation}
while for the two degrees of freedom $\zeta_1$ and $\zeta_2$ for the intrinsic spin-1 flexion $\Delta \zeta$ one derives
\begin{equation}\label{eq:explicitflex1}
\begin{split}
\frac{4}{\sqrt{3}}\frac{\zeta_1}{\xi} = & \frac{1}{\sqrt{3}\, \xi} \Delta q_{abc} \Delta^{(1)}_{abc} = \frac{1}{\sqrt{3} \xi} \int \mathrm{d}^2 r \rho(r) \frac{\Phi_{efg}}{6 \sigma^2} r_e r_f r_g  r_a r_b r_c \Delta^{(1)}_{abc} = \frac{1}{\sqrt{3} \xi} \int \mathrm{d}^2 r \rho(r) \frac{\Phi_{efg}}{6 \sigma^2} r_e r_f r_g\sqrt{3}\left(r_0^3+r_0 r_1^2\right)
\
\\
= &\frac{\int \mathrm{d} r \rho(r) r^7 }{  2 \uppi \int \mathrm{d} r \rho(r) r^5 \cdot 6 \sigma^2} \int_0^{2 \uppi} \mathrm{d} \phi  \left(\cos^3\left(\phi\right)+\cos\left(\phi\right) \sin^2\left(\phi\right)\right)\left( \Phi_{000} \cos^3\left(\phi\right) + 3 \Phi_{001} \cos^2\left(\phi\right) \sin \left(\phi\right) + 3 \Phi_{011} \cos\left(\phi\right) \sin^2\left(\phi\right) + \Phi_{111} \sin^3\left(\phi\right)\right) 
\
\\
= & \frac{3} {8} \frac{1}{6 \sigma^2} \tilde{D}_\mathcal{F}  \partial_0 \Delta \Phi,
\end{split}
\end{equation}
and, in a similar calculation, 
\begin{equation}
\frac{4}{\sqrt{3}}\frac{\zeta_2}{\xi} = \frac{3}{8} \frac{1}{6 \sigma^2} \tilde{D}_\mathcal{F} \partial_1 \Delta \Phi\,,
\end{equation}
such that the intrinsic spin-1 flexion $\Delta \zeta$ is just the gradient of the intrinsic size variation $\partial \left(\Delta s\right)$,
\begin{equation}\label{eq:Deltazeta}
\Delta \zeta = \frac{4}{\sqrt{3}} \frac{\zeta_1 + \mathrm{i} \zeta_2}{\xi} = \frac{3}{8} \frac{1}{6 \sigma^2} \tilde{D}_\mathcal{F}  \left(\partial_0 + \mathrm{i} \partial_1\right) \Delta \Phi = \frac{1}{2} \frac{\tilde{D}_\mathcal{F}}{\tilde{D}_\epsilon} \partial\left(\Delta s \right) = 3 \frac{D_{IA,\mathcal{F}}}{D_{IA}} \partial \left(\Delta s\right)\varpropto  \partial \left(\Delta s\right) \,.
\end{equation}
This result is analogous to the weak lensing case, where the spin-1 flexion $\mathcal{F}$ can just be expressed as the gradient of the convergence $\kappa$ as $\mathcal{F}=\partial \kappa$ as defined in \citet{bacon_weak_2006}. Accordingly, one can continue to relate the variations of the intrinsic complex ellipticity $\Delta \epsilon = \epsilon/s_0$
\begin{equation}
\Delta \epsilon = 
\frac{\epsilon_{+} + \mathrm{i} \epsilon_{\times}}{s_0} = \frac{\Delta q_{cd} \sigma^{(3)}_{cd} + \mathrm{i} \Delta q_{cd} \sigma^{(0)}_{cd}}{2 s_0} = \frac{1}{8 \sigma^2} \tilde{D}_\epsilon \left(\frac{1}{2}\left(\Phi_{00} - \Phi_{11}\right) + \mathrm{i} \Phi_{01} \right)
\end{equation}
to the intrinsic spin-3 flexion $\Delta \delta$, for which one explicitly obtains
\begin{equation}\label{eq:Deltadelta}
\Delta \delta = 
4 \frac{\delta_1 + \mathrm{i} \delta_2}{\xi} = \frac{\Delta q_{abc} \Delta^{(3)}_{abc} + \mathrm{i} \Delta q_{abc} \Delta^{(4)}_{abc}}{\xi} =\frac{1} {8 \cdot 6 \sigma^2} \tilde{D}_\mathcal{F}  \left[\left( \Phi_{000} - 3 \Phi_{011} \right) + \mathrm{i} \left(3 \Phi_{001} - \Phi_{111} \right)\right] = \frac{1}{3} \frac{ \tilde{D}_\mathcal{F} }{\tilde{D}_\epsilon} \partial\left(\Delta \epsilon \right) = 2  \frac{ D_{IA,\mathcal{F}}}{D_{IA}} \partial\left(\Delta \epsilon \right) \varpropto  \partial\left(\Delta \epsilon \right).
\end{equation}
Again, one immediately recognises the analogy to the weak lensing result, where the spin-3 flexion $\mathcal{G}$ from \citet{bacon_weak_2006} is just the gradient of the shear $\gamma$, i.e. $\mathcal{G} = \partial \gamma$. The results (\ref{eq:Deltazeta}) and (\ref{eq:Deltadelta}) are important for they in principle show that when performing an analysis of the intrinsic flexions the power spectra of intrinsic flexions and intrinsic ellipticities are in principle sourced by the same alignment potential $\varphi$ as a line-of sight average over the physical gravitational potential, weighted by the redshift distribution 
\begin{equation}
p(z)\varpropto \left( \frac{z}{z_0} \right)^2 \exp \left[- \left( \frac{z}{z_0} \right)^\beta \right],
\end{equation} 
with parameters $\beta = 3/2$ and $z_0 = 0.64$ typically adopted in studies for the Euclid-mission \citep{laureijs2011euclid}. It has been shown that different tomographic binning strategies do not have a major impact on the precision of the derived parameters in a standard cosmology \citet{2019arXiv190106495K, Sipp}, and we work for simplicity with equally populated bins where the Poissonian contribution shape noise is constant.

The first step hereby is to express the intrinsic flexions in terms of angular coordinates $\theta$, where $r = \theta \chi$ is the transverse comoving separation of different points, while $\theta$ is the respective angular separation and $\chi$ denotes the comoving distance between the galaxy and the observer, and spatial derivatives transform accordingly as $\partial_{r_a} \equiv \partial_{\theta_a} \chi^{-1}$. For notational simplicity angular derivatives will be denoted as $\partial_{\theta_a}\equiv \partial_a$ from now on. Thus, the flexions in angular coordinates and the flexions in spatial coordinates are related by
\begin{equation}\label{eq:conversion}
\left(\Delta \zeta\right)_{(r)}, \left(\Delta \delta\right)_{(r)} \varpropto 
\frac{\Delta q_{abd}}{q_{abcd}} \varpropto \frac{ \int \mathrm{d}^2 r \rho(r) r_a r_b r_c}{\int \mathrm{d}^2 r \rho_0(r) r_a r_b r_c r_d} \varpropto \frac{\chi^3}{\chi^4} \frac{ \int \mathrm{d}^2 \theta \rho(\theta) \theta_a \theta_b \theta_c}{\int \mathrm{d}^2 \theta \rho_0(\theta) \theta_a \theta_b \theta_c \theta_d} \varpropto 
\chi^{-1}\left(\Delta \zeta\right)_{(\theta)}, \chi^{-1}\left(\Delta \delta\right)_{(\theta)}\,.
\end{equation} 
The intrinsic alignment potential $\varphi_A$ was introduced by \citet{ghosh2020intrinsic} to be
\begin{equation}
\varphi_A = 
\int_{\chi_\text{A}}^{\chi_\text{A+1}} \mathrm{d} \chi \,  p(\chi) \text{H}(\chi) D_{IA} \, \frac{D_+ (a)}{a} \frac{1}{\chi^2} \frac{\Phi}{c^2} = \int \mathrm{d} \chi W_{\varphi,A}\left(\chi\right) \frac{\Phi}{c^2}, \quad \text{with weighting} \, W_{\varphi,A} = p(\chi) \Theta\left(\chi - \chi_\text{A}\right) \Theta\left(\chi_\text{A+1} - \chi\right) \text{H}(\chi) D_{IA} \, \frac{D_+ (a)}{a} \frac{1}{\chi^2}
\end{equation} 
where the Hubble-function is inserted as $- \mathrm{d} \chi \text{H}(\chi) = c\mathrm{d} z $, and the time evolution of the potential is given by the linear growth factor $D_+ (a)/a$ while $\Theta$ denotes the Heaviside-function defining tomographic bin $A$. Thus, with the alignment potential $\varphi_A$ and the conversion of the flexions from radial to angular coordinates (\ref{eq:conversion}) the tomographic averages of the flexion components in angular units (\ref{eq:zeta1}) to (\ref{eq:delta1}) become:
\begin{align}
\label{eq:deltatomo}
&\frac{\delta_1}{\xi}_A =  
\frac{1}{4} \frac{D_{IA,\mathcal{F}}}{D_{IA}} \partial_a\partial_b\partial_c \varphi_A \Delta^{(3)}_{abc}, \, \, \frac{\delta_2}{\xi}_A  =\frac{1}{4} \frac{D_{IA,\mathcal{F}}}{D_{IA}}\partial_a\partial_b\partial_c \varphi_A \Delta^{(4)}_{abc}, \quad \frac{\zeta_1}{\xi}_A  =\frac{3}{4} \frac{D_{IA,\mathcal{F}}}{D_{IA}} \partial_a\partial_b\partial_c \varphi_A \Delta^{(1)}_{abc}, \, \, \frac{\zeta_2}{\xi}_A  =\frac{3}{4} \frac{D_{IA,\mathcal{F}}}{D_{IA}}\partial_a\partial_b\partial_c \varphi_A  \Delta^{(2)}_{abc},
\end{align}
where the additional factor $3$ for $(\zeta_1/\xi)_A$ and $(\zeta_2/\xi)_A$ takes into account that the distortions due to the intrinsic spin-1 field are thrice as strong as the ones sourced by the intrinsic spin-3 field a shown in (\ref{eq:zeta1}). Consequently, one can decompose the third derivative of the alignment potential accordingly into the basis of matrices (\ref{eq:Diracs}) as
\begin{equation}
\partial_a\partial_b\partial_c \varphi_A= \frac{D_{IA}}{D_{IA,\mathcal{F}}} \left(\frac{1}{3}\frac{\zeta_1}{\xi}_A \Delta^{(1)}_{abc} +\frac{1}{3}\frac{\zeta_2}{\xi}_A \Delta^{(2)}_{abc}+ \frac{\delta_1}{\xi}_A \Delta^{(3)}_{abc}+\frac{\delta_2}{\xi}_A \Delta^{(4)}_{abc}\right).
\end{equation}

For the weak lensing flexion one can perform a similar decomposition of the lensing potential $\psi_B $ averaged over a tomographic redshift bin $B$ \citep{hu_dark_2002, jain_cross-correlation_2003, huterer_nulling_2005, amara_optimal_2007, takada_tomography_2004, munshi_tomography_2014}: The lensing potential with the respective lensing efficiency $W_{\Psi,B}$ function is given as
\begin{equation}
\psi_B = 
\int \mathrm{d} \chi W_{\Psi,B} \left(\chi\right) \frac{\Phi}{c^2}, \quad  W_{\Psi,B} = \frac{2}{\chi} \frac{D_+ (a)}{a} \int_{\text{max}\left(\chi, \chi_B\right)}^{\chi_{B+1}} \mathrm{d} \chi^\prime \, p\left(\chi^\prime\right) \frac{\mathrm{d} z}{\mathrm{d} \chi^\prime} \left(1 - \frac{\chi}{\chi^\prime}\right)\,.
\end{equation}
Therefore, the decomposition of the lensing flexion, as also discussed in Appendix~\ref{appA} is given by 
\begin{equation}
\partial_a \partial_b \partial_c \psi_B =  -\frac{\sqrt{3}}{2} \mathcal{F}_{1,B} \Delta^{(1)}_{abc}-\frac{\sqrt{3}}{2} \mathcal{F}_{2,B} \Delta^{(2)}_{abc}-\frac{1}{2} \mathcal{G}_{1,B} \Delta^{(3)}_{abc}-\frac{1}{2} \mathcal{G}_{2,B} \Delta^{(4)}_{abc}
\end{equation}
where each coefficient can be received via orthogonal projection with the corresponding $\Delta$-matrix as
\begin{align}\label{eq:flextomo}
-\frac{\sqrt{3}}{2} \mathcal{F}_{1,B} = \frac{1}{4} \partial_a \partial_b \partial_c \psi_B \Delta^{(1)}_{abc}, \,\, -\frac{\sqrt{3}}{2} \mathcal{F}_{2,B} = \frac{1}{4} \partial_a \partial_b \partial_c \psi_B \Delta^{(2)}_{abc}\,,
\quad
-\frac{1}{2} \mathcal{G}_{1,B} = \frac{1}{4} \partial_a \partial_b \partial_c \psi_B \Delta^{(3)}_{abc}, \, \, -\frac{1}{2} \mathcal{G}_{2,B} = \frac{1}{4} \partial_a \partial_b \partial_c \psi_B \Delta^{(4)}_{abc}\,,
\end{align}
The intrinsic flexions can then be correlated with the weak lensing flexions to find the observability of the intrinsic flexion compared to the lensing flexion, as discussed in the next section.





\section{Angular spectra of intrinsic and extrinsic flexions}\label{sect_spectra}
In order to derive the power spectra for intrinsic flexions as well as the according cross correlations with the lensing flexions we assume the statistical fluctuations in the density variations to be described by homogeneous and isotropic Gaussian random fields on the sphere and apply the Limber-approximation for relating the three-dimensional correlations of the source fields to the two-dimensional, angular spectra of the observables \citep{limber}. Then, the intrinsic self-correlation, i.e. the $II$-correlation resulting from the alignment potential $\varphi_A$ becomes
\begin{equation}\label{eq:IIcorrelations}
\begin{split}
\left\langle \varphi_{A,abc}\left(\ell\right) \varphi^*_{B,def}\left(\ell^\prime\right) \right\rangle = &\left(2 \uppi\right)^2 \delta_D\left(\ell - \ell^\prime\right) C^{\varphi_A \varphi_B}_{abcdef}\,,
\
\\
\text{with} \quad C^{\varphi_A \varphi_B}_{abcdef} = &\ell_a \ell_b \ell_c \ell_d \ell_e \ell_f \left\langle \int \mathrm{d} \chi W_{\varphi_A} \frac{\Phi}{c^2}\left( k=\ell/\chi, \chi \right) \int \mathrm{d} \chi' W_{\varphi_B} \frac{\Phi}{c^2}\left( k'=\ell'/\chi', \chi' \right) \right\rangle 
\
\\
= & \ell_a \ell_b \ell_c \ell_d \ell_e \ell_f  \int \mathrm{d} \chi \frac{W_{\varphi_A} W_{\varphi_B}}{\chi^2} P_{\Phi \Phi}\left( k=\ell/\chi, \chi \right) = \ell_a \ell_b \ell_c \ell_d \ell_e \ell_f  C^{\varphi_A \varphi_B} (\ell)\,,
\end{split}
\end{equation}\citep{Bardeen1986}, 
where the angular derivatives become the reciprocal wave vector modes and the short hand notation $\varphi_{A,abc}$ denotes $\partial_a \partial_b \partial_c \varphi_{A}$. Here, $P_{\Phi \Phi} \varpropto k^{n_s - 4} T(k)^2$ is the power spectrum of the potential fluctuations sourced from the fluctuations in the cosmic density field with spectral index $n_s \leq 1$. $T(k)$ denotes the transfer function which is extended to non-linear scales \citep{Smith2003}. As in the previous work by \citet{ghosh2020intrinsic} we suppress the amplitude of the potential $\Phi$ on small scales with a Gaussian filter, and this scale is chosen to correspond the size of a typical elliptical galaxy.
 
Similarly, for the $GI$-correlations we obtain
\begin{equation}
\left\langle \psi_{A,abc}\left(\ell\right) \varphi^*_{B,def}\left(\ell^\prime\right) \right\rangle = \left(2 \uppi\right)^2 \delta_D\left(\ell - \ell^\prime\right) C^{\psi_A \varphi_B}_{abcdef}\,, \quad \text{with} \, C^{\psi_A \varphi_B}_{abcdef} = \ell_a \ell_b \ell_c \ell_d \ell_e \ell_f  \int \mathrm{d} \chi \frac{W_{\psi_A} W_{\varphi_B}}{\chi^2} P_{\Phi \Phi}\left( k=\ell/\chi, \chi \right) = \ell_a \ell_b \ell_c \ell_d \ell_e \ell_f  C^{\psi_A \varphi_B} (\ell)\,,
\end{equation}
while the autocorrelation of the lensing potential, i.e. the $GG$-term, is given by
\begin{equation}\label{eq:GGcorrelation}
\left\langle \psi_{A,abc}\left(\ell\right) \psi^*_{B,def}\left(\ell^\prime\right) \right\rangle = \left(2 \uppi\right)^2 \delta_D\left(\ell - \ell^\prime\right) C^{\psi_A \psi_B}_{abcdef}, \quad
\text{with} \, C^{\psi_A \psi_B}_{abcdef} = \ell_a \ell_b \ell_c \ell_d \ell_e \ell_f  \int \mathrm{d} \chi \frac{W_{\psi_A} W_{\psi_B}}{\chi^2} P_{\Phi \Phi}\left( k=\ell/\chi, \chi \right) = \ell_a \ell_b \ell_c \ell_d \ell_e \ell_f  C^{\psi_A \psi_B} (\ell)\,.
\end{equation}

For deriving the correlations of the flexions from the correlators of the potentials (\ref{eq:IIcorrelations}) to (\ref{eq:GGcorrelation}) we consider the following orthonormal decomposition of  $\ell_a \ell_b \ell_c$ in terms of the $\Delta$-matrices (\ref{eq:Diracs}) to simplify the according contractions as
\begin{equation}\label{eq:Ldecomposition}
\ell_a \ell_b \ell_c = \frac{1}{4} \left[\sqrt{3} \ell^3 \cos \phi \Delta^{(1)}_{abc} + \sqrt{3} \ell^3 \sin \phi \Delta^{(2)}_{abc} + \ell^3 \cos \left( 3 \phi \right)  \Delta^{(3)}_{abc} + \ell^3 \sin \left( 3 \phi \right)  \Delta^{(4)}_{abc} \right] = \frac{1}{4} \left[\sqrt{3} \ell^3 \Delta^{(1)}_{abc} + \ell^3 \Delta^{(3)}_{abc} \right]\, \, \, \text{for $\phi=0$},
\end{equation}
with $\ell_0 = \ell \cos \phi$ and $\ell_1 = \ell \sin \phi$. Isotropy of the random field allows to set $\phi=0$ for evaluation of the terms. The result (\ref{eq:Ldecomposition}) also confirms the different spin symmetries of the flexion fields, since the contractions of the matrices $\Delta^{(1)}$ and $\Delta^{(2)}$ with $\ell_a \ell_b \ell_c$ are of spin-1 while the  contractions with the other two matrices $\Delta^{(3)}$ and $\Delta^{(4)}$ have spin-3. Thus, the intrinsic spin-1 flexion spectrum $C^{ \Delta \zeta^\prime \Delta \zeta^\prime}$ with $\Delta \zeta^\prime = 4/9 \, \Delta \zeta$ can easily be derived from the autocorrelation of the $\zeta_1/\xi$-component as
\begin{equation}
\begin{split}
\text{$II$:}
\quad C^{ \Delta \zeta^\prime  \Delta \zeta^\prime}_{AB}\left( \ell \right)= C^{ \frac{16}{9\sqrt{3}}\frac{\zeta_1}{\xi}  \frac{16}{9\sqrt{3}}\frac{\zeta_1}{\xi}}_{AB}\left( \ell \right)=\frac{16}{27} \frac{D_{IA,\mathcal{F}}^2}{D_{IA}^2} \Delta^{(1)}_{abc} \Delta^{(1)}_{def}  \ell_a \ell_b \ell_c \ell_d \ell_e \ell_f C^{\varphi_A \varphi_B} (\ell) = \frac{16}{9} \frac{D_{IA,\mathcal{F}}^2}{D_{IA}^2} \ell^6  C^{\varphi_A \varphi_B} (\ell),
\end{split}
\end{equation}
where the additional factor $4/\sqrt{3}$ comes from relation (\ref{eq:spin1holic}). Furthermore, weighting the intrinsic spin-1 flexion field $\Delta \zeta$ with the prefactor $4/9$ one directly relates the magnitude of the intrinsic flexion field to the lensing flexion field $\mathcal{F}$ due to the approximate relations between the observable HOLICs $\left\langle \zeta \right\rangle$ and $\left\langle \delta \right\rangle$ respectively, the lensing flexions $\mathcal{F}$ and $\mathcal{G}$ respectively, and finally the intrinsic HOLICs or flexions $\left\langle \zeta^s \right\rangle$ and $\left\langle \delta^s \right\rangle$ respectively derived by \citet{okura_new_2007}:
\begin{align}
\label{eq:flexhol}
\left\langle \zeta \right\rangle \approx \left\langle \zeta^s \right\rangle + \frac{9}{4} \mathcal{F}, \quad \text{and} \,\, \left\langle \delta \right\rangle \approx \left\langle \delta^s \right\rangle + \frac{3}{4} \mathcal{G}\,.
\end{align}
Analogously, the power spectrum for the intrinsic spin-3 flexion $\Delta \delta^\prime= 4/3 \Delta \delta$ due to relation (\ref{eq:flexhol}) with additional weighting factor $4$ due to relation~(\ref{eq:spin3holic}) is given by
\begin{equation}
\begin{split}
C^{ \Delta \delta^\prime  \Delta \delta^\prime}_{AB}\left( \ell \right) = 
\frac{16}{9} \frac{D_{IA,\mathcal{F}}^2}{D_{IA}^2} \ell^6  C^{\varphi_A \varphi_B} (\ell), 
\end{split}
\end{equation}
while the cross correlation between the intrinsic spin-1 flexion $\Delta \zeta^\prime$ and the intrinsic spin-3 flexion spectrum $\Delta \delta^\prime$ becomes
\begin{equation}
\begin{split}
C^{ \Delta \zeta^\prime  \Delta \delta^\prime}_{AB}\left( \ell \right) = 
\frac{16}{9} \frac{D_{IA,\mathcal{F}}^2}{D_{IA}^2} \ell^6  C^{\varphi_A \varphi_B} (\ell) \,.
\end{split}
\end{equation}

In the course of this work we only want to consider correlations of the intrinsic spin-1 flexion and the according lensing flexion $\mathcal{F}$, since the noise of the spin-3 flexion is usually three times higher than the noise of the spin-1 flexion as discussed in \citet{okura_new_2007} and would yield only detections at correspondingly lower significance. As will be discussed below and in Sect.~\ref{sect_s2n} the uncertainty for the spin-1 flexion exceeds already several orders of magnitude compared to the actual signal, and only in a cumulated signal-to-noise measurement would it be possible to measure $C^{ \Delta \zeta^\prime  \Delta \zeta^\prime}_{AB}\left( \ell \right)$ or $C^{\mathcal{F} \, \Delta \zeta^\prime}_{AB}\left( \ell \right)$. Thus, we will from now on only state the relevant results for the spin-1 flexion spectra as
\begin{align}
&\text{$GG$:}
\quad C^{\mathcal{F}\mathcal{F}}_{AB}\left( \ell \right)= \frac{1}{16} \left(\frac{-2}{\sqrt{3}}\right)^2 \Delta^{(1)}_{abc} \Delta^{(1)}_{def}  \ell_a \ell_b \ell_c \ell_d \ell_e \ell_f  C^{\psi_A \psi_B} (\ell) =  \frac{1}{4} \ell^6 C^{\psi_A \psi_B} (\ell)\,,
\
\\
&\text{$GI$:}
\quad C^{\mathcal{F} \, \Delta \zeta^\prime}_{AB}\left( \ell \right)= \frac{1}{16} \left(\frac{-2}{\sqrt{3}}\right) \frac{4}{9}\frac{4}{\sqrt{3}} 3 \frac{D_{IA,\mathcal{F}}}{D_{IA}} \Delta^{(1)}_{abc} \Delta^{(1)}_{def}  \ell_a \ell_b \ell_c \ell_d \ell_e \ell_f  C^{\psi_A \varphi_B} (\ell) = -\frac{2}{3}\frac{D_{IA,\mathcal{F}}}{D_{IA}} \ell^6 C^{\psi_A \varphi_B} (\ell)\,.
\end{align}

\begin{figure}
\centering
\includegraphics[scale=0.45]{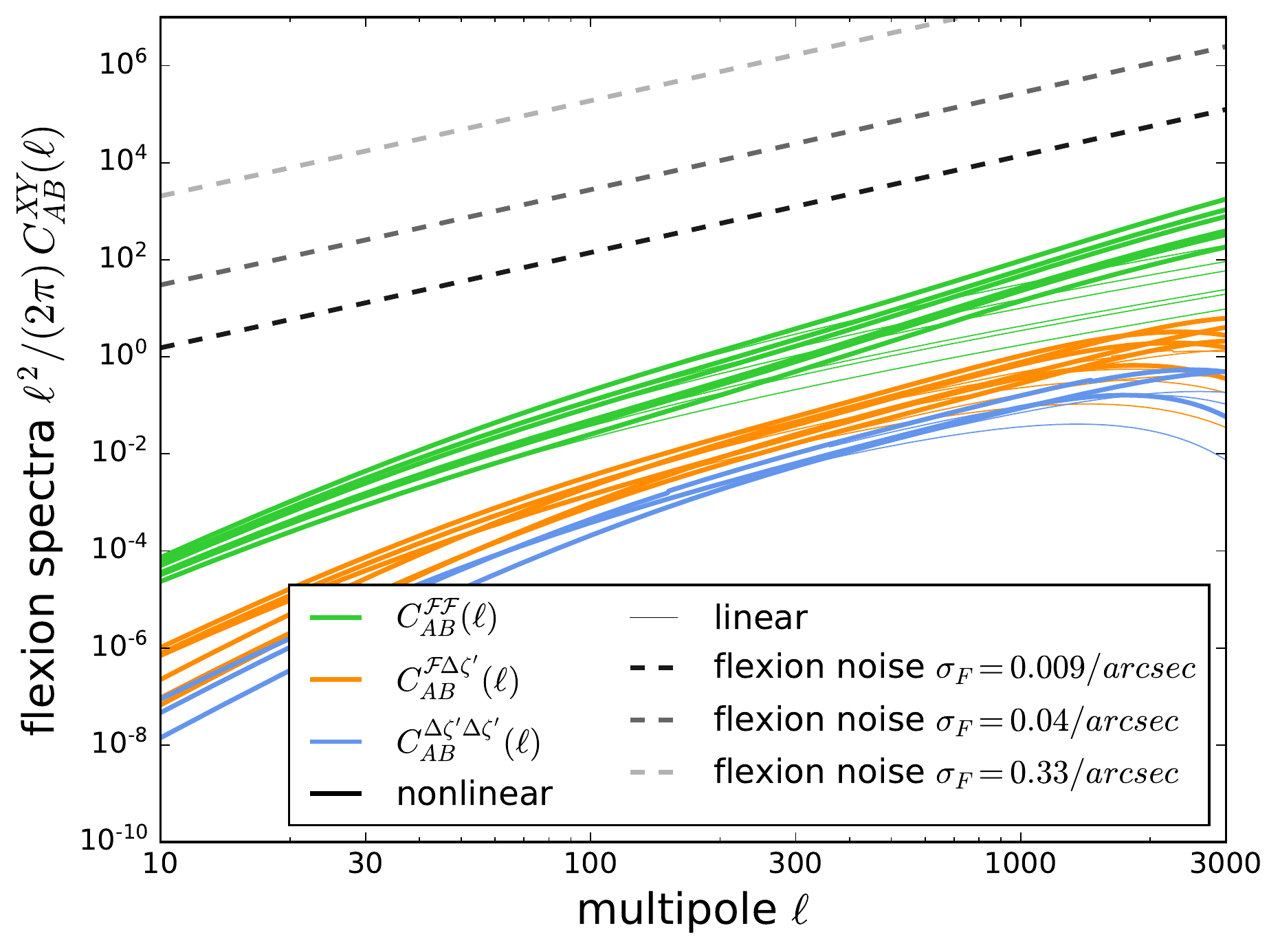}
\caption{Tomographic flexion spectra as a function of multipole order $\ell$, separated by gravitational lensing $C_{AB}^{\mathcal{F}\mathcal{F}}(\ell)$, intrinsic size correlations $C_{AB}^{\Delta\zeta^\prime\Delta\zeta^\prime}(\ell)$ and the cross-correlation $C_{AB}^{\mathcal{F}\Delta\zeta^\prime}(\ell)$ (of which we show the absolute value), with the Poissonian noise contributions $N_{AB}^\mathcal{F}(\ell)$ based on the three flexion dispersions of $\sigma_F= 0.33 \mathrm{arcsec}^{-1}$, $\sigma_F= 0.04 \mathrm{arcsec}^{-1}$ and $\sigma_F= 0.009 \mathrm{arcsec}^{-1}$ (dashed grey lines in different shades) in comparison, for Euclid's redshift distribution and tomography with 3 bins, for a $\Lambda$CDM-cosmology with an alignment parameter $D=-10^{-2}$ on a mass scale $M = 10^{12}M_\odot/h$, corresponding to a virial velocity of $\sigma\simeq10^5\mathrm{m}/\mathrm{s}$. Thick and thin lines indicate a nonlinear and linear spectrum, respectively, and the S{\'e}rsic-index was set to $n=1$, corresponding to the exponential profile. There are six thick respectively thin lines for the six possible redshift bin combinations for $C_{AB}^{\mathcal{F}\mathcal{F}}(\ell)$ and $C_{AB}^{\mathcal{F}\Delta\zeta^\prime}(\ell)$, while the intrinsic size correlation $C_{AB}^{\Delta\zeta^\prime\Delta\zeta^\prime}(\ell)$ is diagonal such that there are only three different lines for either linear or non-linear spectra for low, intermediate and high redshift bins.}
\label{fig:shapeshape_1}
\end{figure}

Fig.~\ref{fig:shapeshape_1} illustrates the according auto-spectra $C^{\mathcal{F}\mathcal{F}}_{AB}(\ell)$, $C^{\mathcal{F}\Delta\zeta^\prime}_{AB}(\ell)$ and the absolute value of the cross-spectra $C^{\Delta\zeta^\prime\Delta\zeta^\prime}_{AB}(\ell)$, which is only nonzero for $A = B$, i.e. within the same bin, due to the locality of intrinsic alignments. As flexions, irrespective of being intrinsic or extrinsic, are generated by third derivatives of the gravitational potential they differ by a factor of $\ell^2$ from ellipticity correlations, which reflect the second derivatives of the gravitational potential. Consequently, one is dealing with very steep spectra, increasing dramatically as a function of multipole $\ell$. For values of the alignment parameter $D_{IA}$ consistent with intrinsic ellipticity correlations and for a low value of the S{\'e}rsic-index $n$ one recovers the same ordering of the effects: $GG$-terms are largest, followed by $GI$- and lastly, $II$-terms. The poissonian noise spectra $N_{AB}(\ell) = \sigma^2_{\mathcal{F}} n_{\text{tomo}}/\bar{n} \delta_{AB}$ are constant and the value for $n_{\text{tomo}}$ denotes the number of tomographic bins, and one chooses $\bar{n}=3.545\times 10^8 \text{sr}^{-1} = 30 \, \text{arcmin}^{-2}$ as the baseline value of the average number density for surveys like Euclid (\cite{laureijs2011euclid}). This value is less than the too optimistic stretch goal value of $\bar{n}=40 \, \text{arcmin}^{-2}$ selected in the previous study by \citet{ghosh2020intrinsic}. The choice of the dispersion $\sigma_{\mathcal{F}}$ is not straightforward: While \citet{okura_new_2007} suggested a highly optimistic value of $\sigma_{\mathcal{F}} = 0.009 \, \text{arcsec}^{-1}$ per galaxy, more recent studies of galaxy shape measurements from simulations using the \textit{Hubble Space Telescope} Ultra Deep Field data (HUDF) show that the determination of the flexion as well as its uncertainty from pixel data comes with significant challenges \citep{rowe2013}, due to finite photon numbers received by the detectors as well as deblending. Furthermore, the authors \citet{rowe2013} also discuss challenges to identify systematic biases from centroid shifts, the dimensionality of the flexion and its effect on the measurement uncertainty, point spread functions of measurement devices, and methods to calibrate flexion measurements using a shapelet decompostion, all of which strictly limit to measure even lensing flexions from realistic data. Especially, \citet{rowe2013} demonstrated that the measured flexion uncertainity depends on the signal to noise ratio (SNR) of the galaxies via a steeply increasing power law: While in the limit of very high signal to noise ratios ($\mathrm{SNR}\approx 1300$) the optimistic dispersion value of $\sigma_{\mathcal{F}} = 0.04 \, \text{arcsec}^{-1}$ from \cite{goldberg_galaxy-galaxy_2005} could be considered as good estimate, the uncertainty of  $\sigma_{\mathcal{F}} = 0.33 \, \text{arcsec}^{-1}$ was more accurate for middle range signal to noise ratios of $\mathrm{SNR} = 100$ in the simulation, and in the limit of very low signal to noise ratio ($\mathrm{SNR}=10$) the dispersion was determined to be about $\sigma_{\mathcal{F}} \approx 2 \, \text{arcsec}^{-1}$. For the value by \citet{okura_new_2007} of $\sigma_{\mathcal{F}} = 0.009 \, \text{arcsec}^{-1}$ the signal to noise ratio of the measured galaxies would even need to be around $\mathrm{SNR} \approx 7700$ according to the power law suggested by \cite{rowe2013}. In this work the poissonian noise spectra, as well as the cumulated signal to noise ratios should be derived for both the optimisitc value of $\sigma_{\mathcal{F}} = 0.04 \, \text{arcsec}^{-1}$ as well as the more realistic value of $\sigma_{\mathcal{F}} = 0.33 \, \text{arcsec}^{-1}$ from \cite{rowe2013} to compare the possible observability of intrinsic flexions for both choices. We also present what would be expected for the dispersion by \citet{okura_new_2007}, however noting that such an optimistic value might not be reachable in pratice, what strongly restricts the possibility to measure intrinsic flexions. Due to the low value for the flexion spectra in comparison to the noise one should expect low values for the attainable signal to noise ratio in an actual measurement, as the measurement at every $\ell$ is noise dominated by about two to orders of magnitude for $\sigma_{\mathcal{F}} = 0.009 \, \text{arcsec}^{-1}$, three to four orders of magnitude for $\sigma_{\mathcal{F}} = 0.04 \, \text{arcsec}^{-1}$, and finally six to seven orders of magnitude for $\sigma_{\mathcal{F}} = 0.33 \, \text{arcsec}^{-1}$, requiring a large range of multipoles to accumulate a signal exceeding the statistical detection threshold.

\begin{figure}
\centering
\includegraphics[scale=0.45]{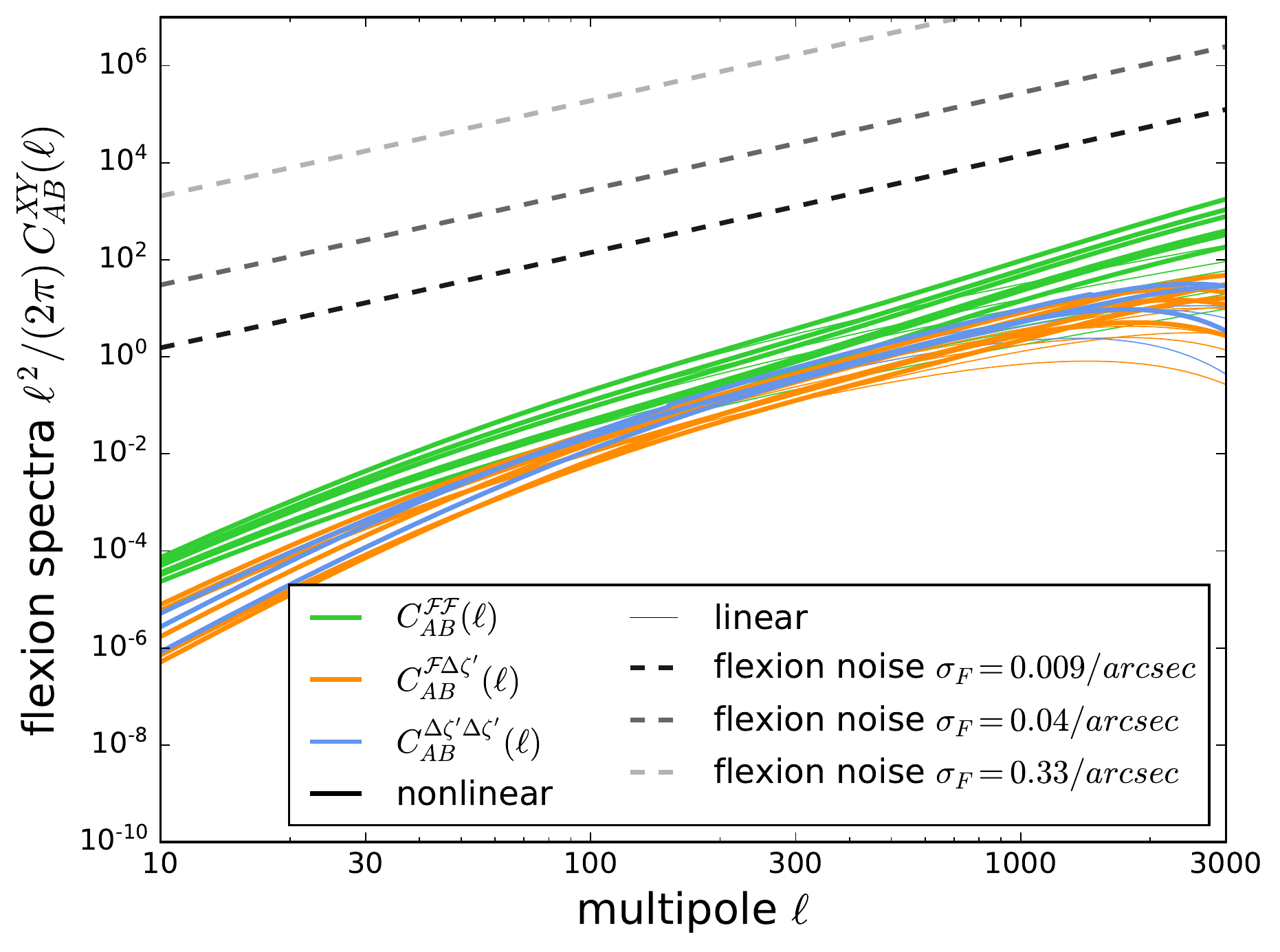}
\caption{Tomographic flexion spectra, similar to the spectra depicted in figure \ref{fig:shapeshape_1}, however with the S{\'e}rsic-index set to $n=4$.}
\label{fig:shapeshape_4}
\end{figure}

Fig.~\ref{fig:shapeshape_4} shows the same result as Fig.~\ref{fig:shapeshape_1}, only for the choice of $n=4$ for the S{\'e}rsic-index instead of $n=1$, i.e. for a de Vaucouleurs-profile instead of an exponential profile. The assumption of a different value for $n$ affects the prefactor $D_{IA,\mathcal{F}}/D_{IA}$ in (\ref{eq:prefactor}) of the $\Delta\zeta^\prime$ field and changes the relative magnitude of the $GI$- and $II$-terms, while the $GG$-spectra remain unchanged. As  a higher value of $n$ implies a larger alignment parameter, the spectra involving intrinsic flexions gain in amplitude; linearly for $GI$- and quadratically for $II$-terms. In an actual observation, it would be reasonable to work with an average value for the S{\'e}rsic-index and hence for the alignment parameter; we would like to show with the values of $n=1$ and $n=4$ two extreme choices.

\begin{figure}
\centering
\includegraphics[scale=0.45]{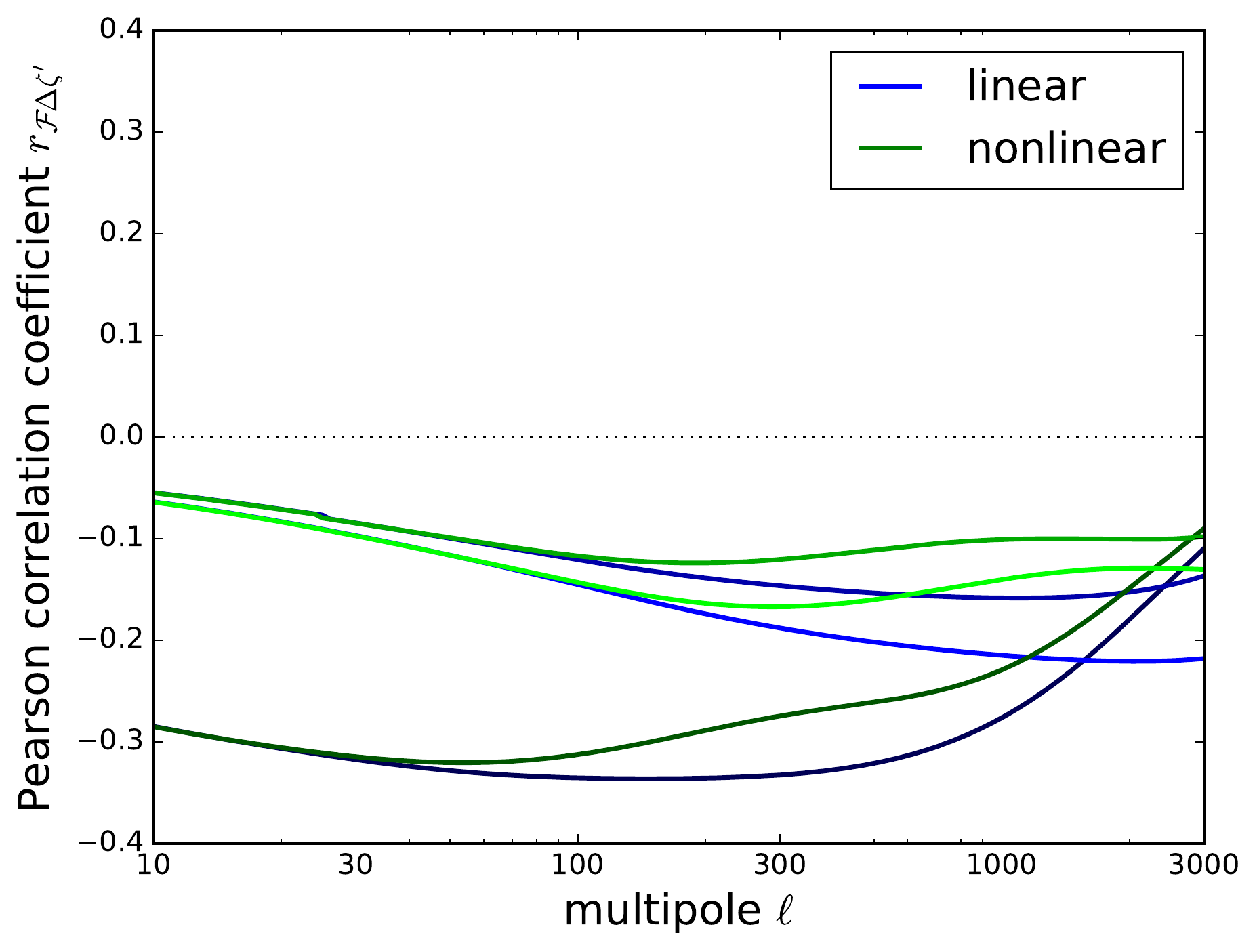}
\caption{Pearson correlation coefficients $r_{\mathcal{F}\Delta\zeta^\prime}(\ell)$ as a function of multipole order $\ell$, derived from a linear and a nonlinear spectrum $P_{\Phi \Phi}(k)$.}
\label{fig:pearson}
\end{figure}

Fig.~\ref{fig:pearson} shows the Pearson correlation coefficient $r_{\mathcal{F}\Delta\zeta^\prime}(\ell)$ as a function of multipole for the intrinsic and extrinsic flexion spectra, defined through
\begin{equation}
r_{\mathcal{F}\Delta\zeta^\prime}(\ell) =
\frac{C^{\mathcal{F}\Delta\zeta^\prime}_{AA}(\ell)}{\sqrt{C^{\mathcal{F}\mathcal{F}}_{AA}(\ell)C^{\Delta\zeta^\prime\Delta\zeta^\prime}_{AA}(\ell)}} = 
- \frac{ -\frac{2}{3}\frac{D_{IA,\mathcal{F}}}{D_{IA}} l^6 C^{\psi_A \varphi_A} (\ell)}{\sqrt{\frac{1}{4} \ell^6 C^{\psi_A \psi_A} (\ell) \, \frac{4^2}{9^2} 9 \frac{D_{IA,\mathcal{F}}^2}{D_{IA}^2} \ell^6  C^{\varphi_A \varphi_A} (\ell) }} = - \frac{C^{\psi_A \varphi_A} (\ell)}{\sqrt{C^{\psi_A \psi_A} (\ell) \, C^{\varphi_A \varphi_A} (\ell) }},
\label{eqn_pearson}
\end{equation}
where the two bin indices are set equal, $A=B$, as otherwise the $II$-type spectra $C^{\Delta\zeta^\prime\Delta\zeta^\prime}_{AB}(\ell)$ would not be defined. Clearly, the negative values of $r_{\mathcal{F}\Delta\zeta^\prime}(\ell)$ illustrate the negative cross-correlation between intrinsic and extrinsic flexion, which is explained by exactly the same arguments as the negative cross-correlation between intrinsic and extrinsic ellipticities discussed by \citet{ghosh2020intrinsic}: In case of an overdense region the background galaxies are magnified due to lensing, while the foreground galaxies actually get smaller due to intrinsic alignment. For an underdense region the opposite is the case. The plot would be absolutely identical for ellipticity correlations given by \citet{ghosh2020intrinsic}, as the two are related by common prefactors of $\ell^2$, which drop out in the Pearson-coefficient, yielding $r_{\mathcal{F}\Delta\zeta^\prime}(\ell) = r_{\gamma\epsilon}(\ell)$. In the end the result eqn.~(\ref{eqn_pearson}) shows that the Pearson-coefficient quantifies in both cases, namely for either intrinsic and extrinsic flexions respectively intrinsic and extrinsic ellipticities, is actually the correlation between the lensing potential $\psi_A$ and the alignment potential $\varphi_A$. The relative difference between the predictions for $r_{\mathcal{F}\Delta\zeta^\prime}(\ell)$ from linear and nonlinear spectra $P_{\Phi \Phi}(k)$ are caused by the tendency of lensing spectra to be affected at higher $\ell$, corresponding to small length scales, by nonlinear contributions to $P_{\Phi \Phi}(k)$ in comparison to alignment spectra, such that the normalisation in eqn.~(\ref{eqn_pearson}) is different.

\section{observability of intrinsic and extrinsic flexions}\label{sect_s2n}
The signal to noise-ratio $\Sigma$ of an observation of a particular contribution $C_{AB}(\ell)$ to the flexion field can be computed in the limit of Gaussian fluctuations statistics by using the relation
\begin{equation}
\Sigma^2 = \sum_\ell \frac{2\ell+1}{2} \text{tr}\left(\mathcal{C}^{-1} S \, \mathcal{C}^{-1} S   \right)\,,
\label{eqn_s2n}
\end{equation}
where $S(\ell)$ is the according signal-matrix under consideration, where $S(\ell)$ is given either by $C^{\mathcal{F} \mathcal{F}}_{A B}(\ell)$, $C^{\mathcal{F} \Delta \zeta'}_{A B}(\ell)$ or $C^{\Delta \zeta' \Delta \zeta'}_{A B}(\ell)$, respectively, for our choice of S{\'e}rsic-indices, as well as galaxy selection mode. Effectively, eqn.~(\ref{eqn_s2n}) results from a Fisher-matrix argument, to what statistical uncertainty $\sigma_A$ an unknown amplitude $A$ of the signal can be determined, resulting in the signal to noise-ratio $\Sigma = A/\sigma_A$ \citep{tegmark_1997, hu_weak_1999}.

\begin{figure}
\centering
\includegraphics[scale=0.45]{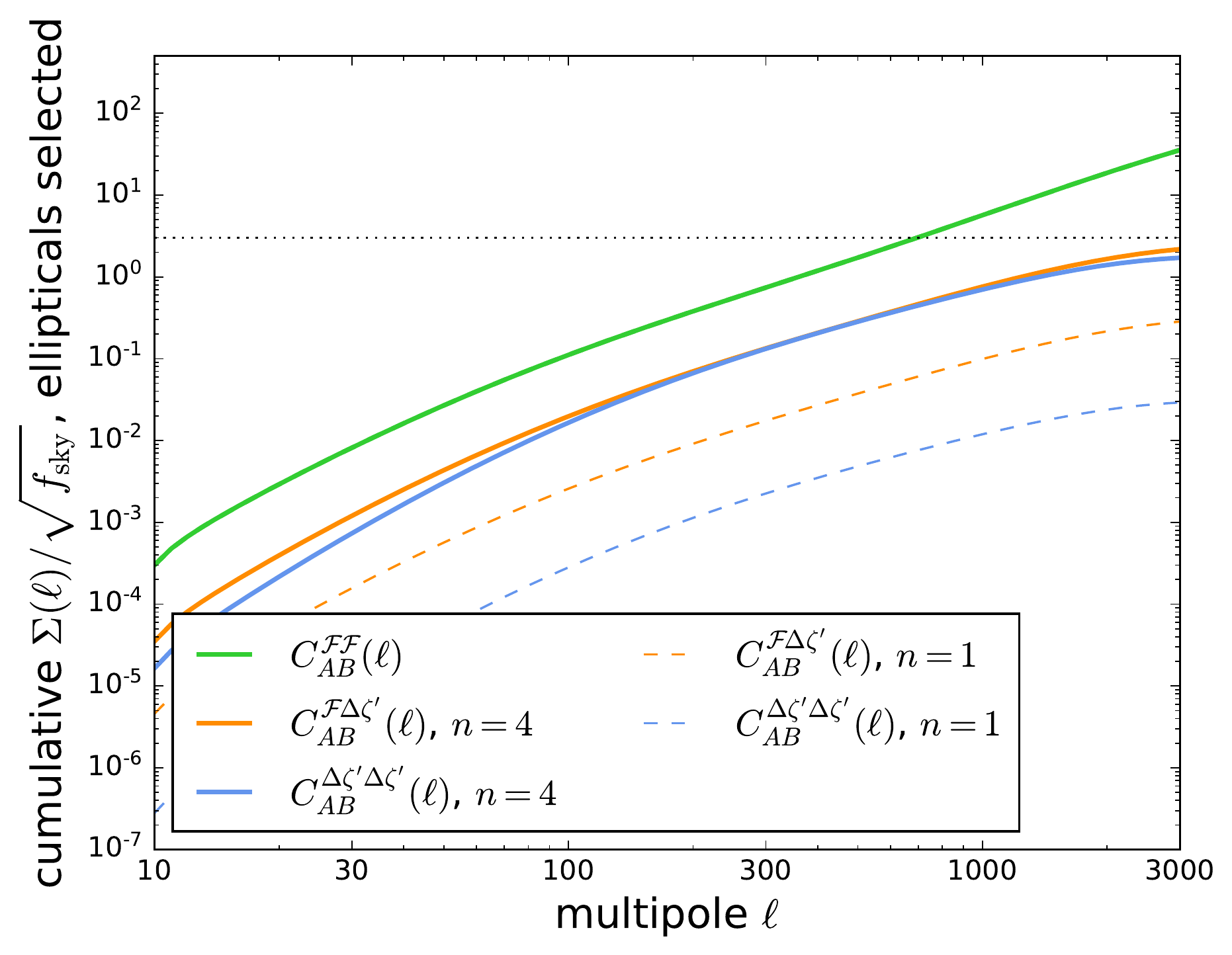}
\caption{Cumulative signal to noise-ratio $\Sigma(\ell)/\sqrt{f_\mathrm{sky}}$ for Euclid 5-bin tomography for measuring intrinsic and extrinsic flexion correlations, for two choices $n=1$ and $n=4$ for the S{\'e}rsic-index, for a sample containing only elliptical galaxies, for a noise level of $\sigma_F=0.009 \text{arcsec}^{-1}$.}
\label{fig:s2n_elliptical}
\end{figure}

\begin{figure}
\centering
\includegraphics[scale=0.45]{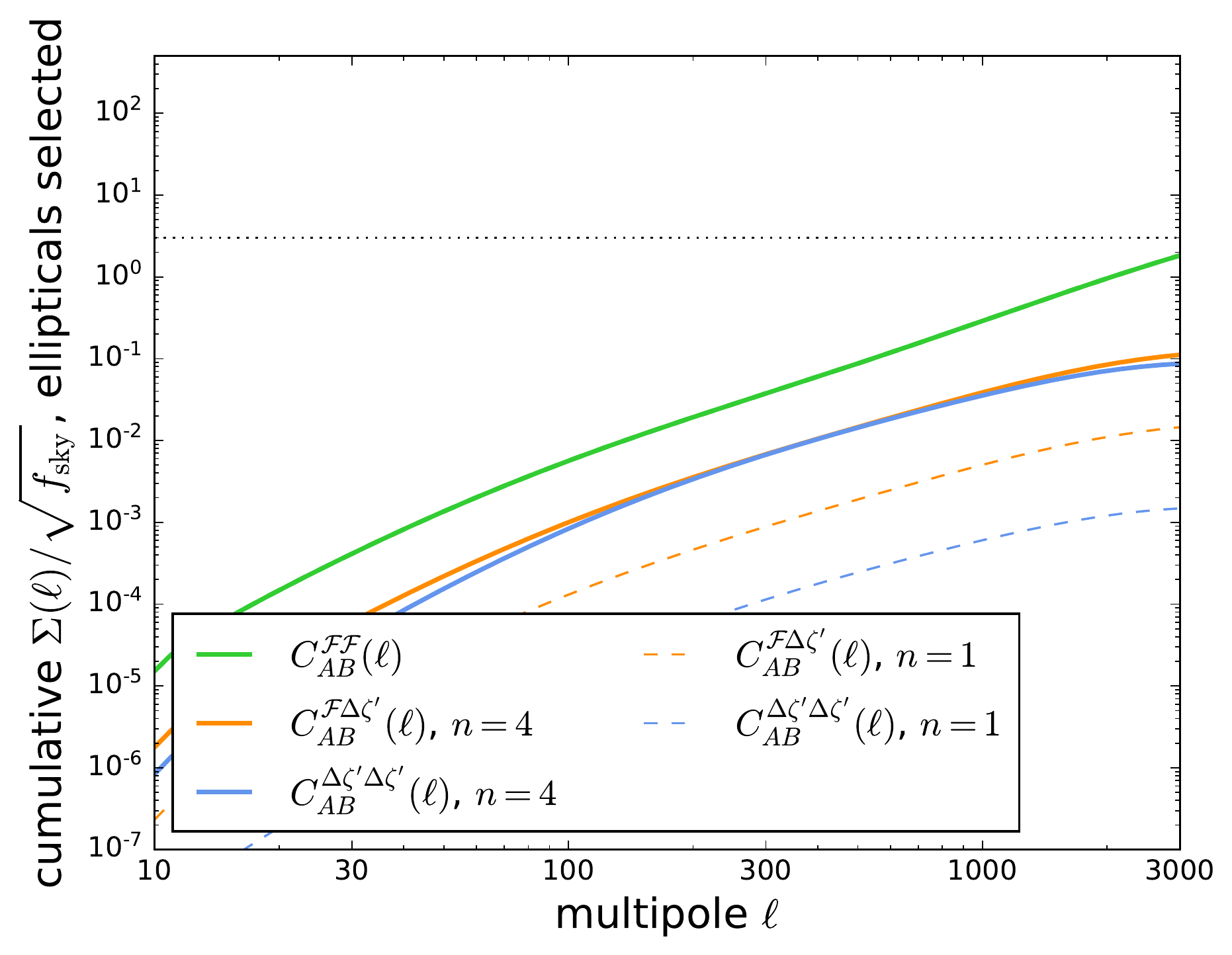}
\caption{Cumulative signal to noise-ratio $\Sigma(\ell)/\sqrt{f_\mathrm{sky}}$ for Euclid 5-bin tomography for measuring intrinsic and extrinsic flexion correlations, for two choices $n=1$ and $n=4$ for the S{\'e}rsic-index, for a sample containing only elliptical galaxies, for a noise level of $\sigma_F=0.04\text{arcsec}^{-1}$.}
\label{fig:s2n_elliptical_Bacon}
\end{figure}

\begin{figure}
\centering
\includegraphics[scale=0.45]{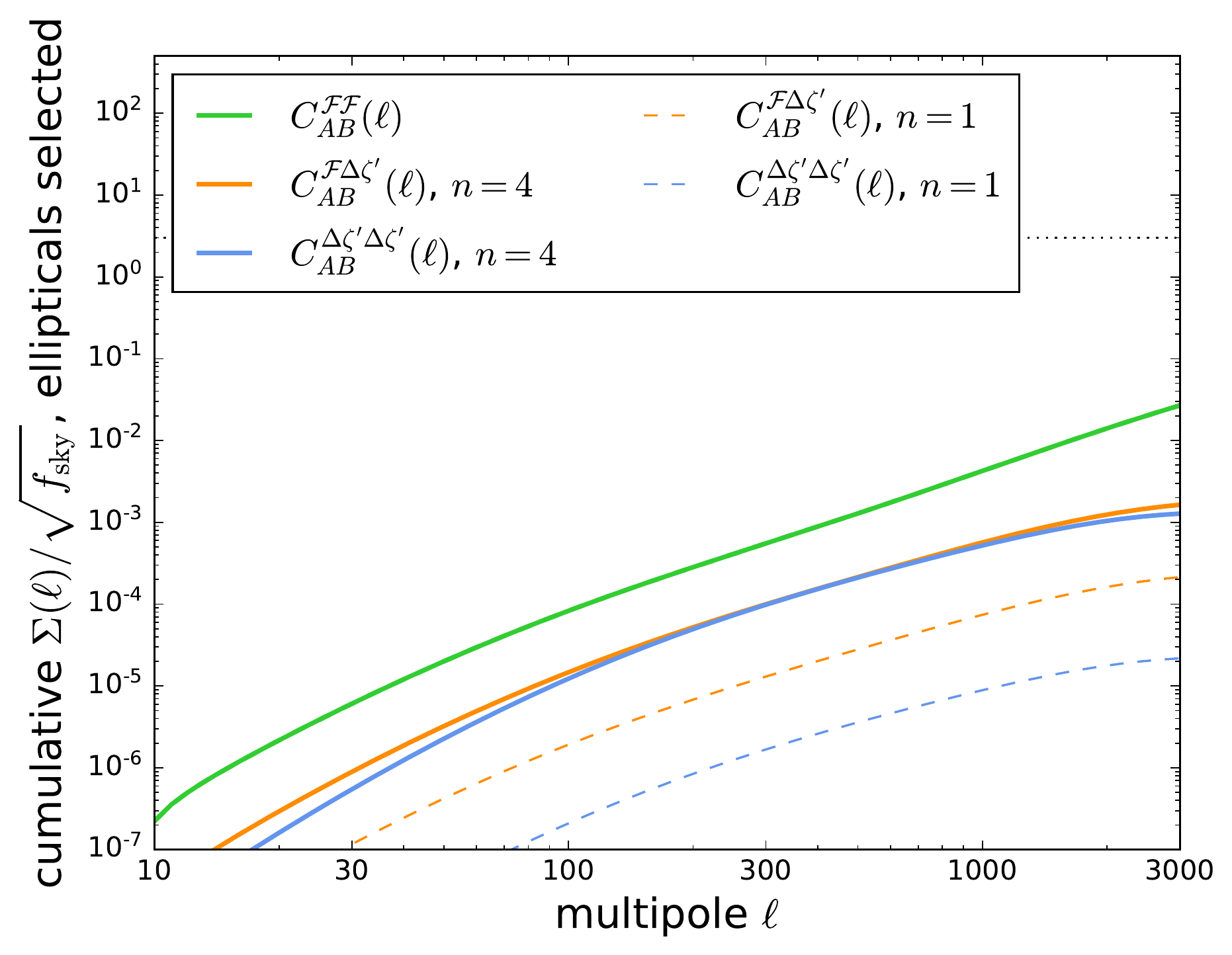}
\caption{Cumulative signal to noise-ratio $\Sigma(\ell)/\sqrt{f_\mathrm{sky}}$ for Euclid 5-bin tomography for measuring intrinsic and extrinsic flexion correlations, for two choices $n=1$ and $n=4$ for the S{\'e}rsic-index, for a sample containing only elliptical galaxies, for a noise level of $\sigma_F=0.33\text{arcsec}^{-1}$.}
\label{fig:s2n_elliptical_Rowe}
\end{figure}

Fig.~\ref{fig:s2n_elliptical} shows the cumulative signal to noise-ratio $\Sigma$ (normalised by $\sqrt{f_\mathrm{sky}}$ to account for incomplete sky coverage).
Working with the highly optimistic noise levels reported by \citet{okura_new_2007} and assuming a S{\'e}rsic-index of $n = 4$ one arrives at the result that a Euclid-like survey would measure $C^{\mathcal{F}\mathcal{F}}_{AB}(\ell)$ at a few ten $\sigma$ of statistical significance, but that $C^{\mathcal{F}\Delta\zeta^\prime}_{AB}(\ell)$ and $C^{\Delta\zeta^\prime\Delta\zeta^\prime}_{AB}(\ell)$ would effectively be hardly detectable at less than $\sim 2\sigma$, even with optimistic assumptions about the noise level and the profile shape, which enters the effective alignment parameter. Lower values of the S{\'e}rsic-index such as $n = 1$ typical for exponential profiles, would yield only very weak intrinsic flexion correlations which are essentially unobservable. However, with the more realistic, yet still optimistic, noise value by \citet{goldberg_galaxy-galaxy_2005} the cumulative signal for elliptical galaxies only as shown in Fig.~\ref{fig:s2n_elliptical_Bacon} is supressed by one order of magnitude such that neither intrinsic nor extrinsic flexions exceed the cumulated signal to noise ratio of $3$ required for observability, while for the realistic dispersion value of $\sigma_F = 0.33 \mathrm{arcsec}^{-1}$ depicted in Fig.~\ref{fig:s2n_elliptical_Rowe} the cumulated signals are even lowered by two to three orders of magnitude compared to Fig.~\ref{fig:s2n_elliptical}.

\begin{figure}
\centering
\includegraphics[scale=0.45]{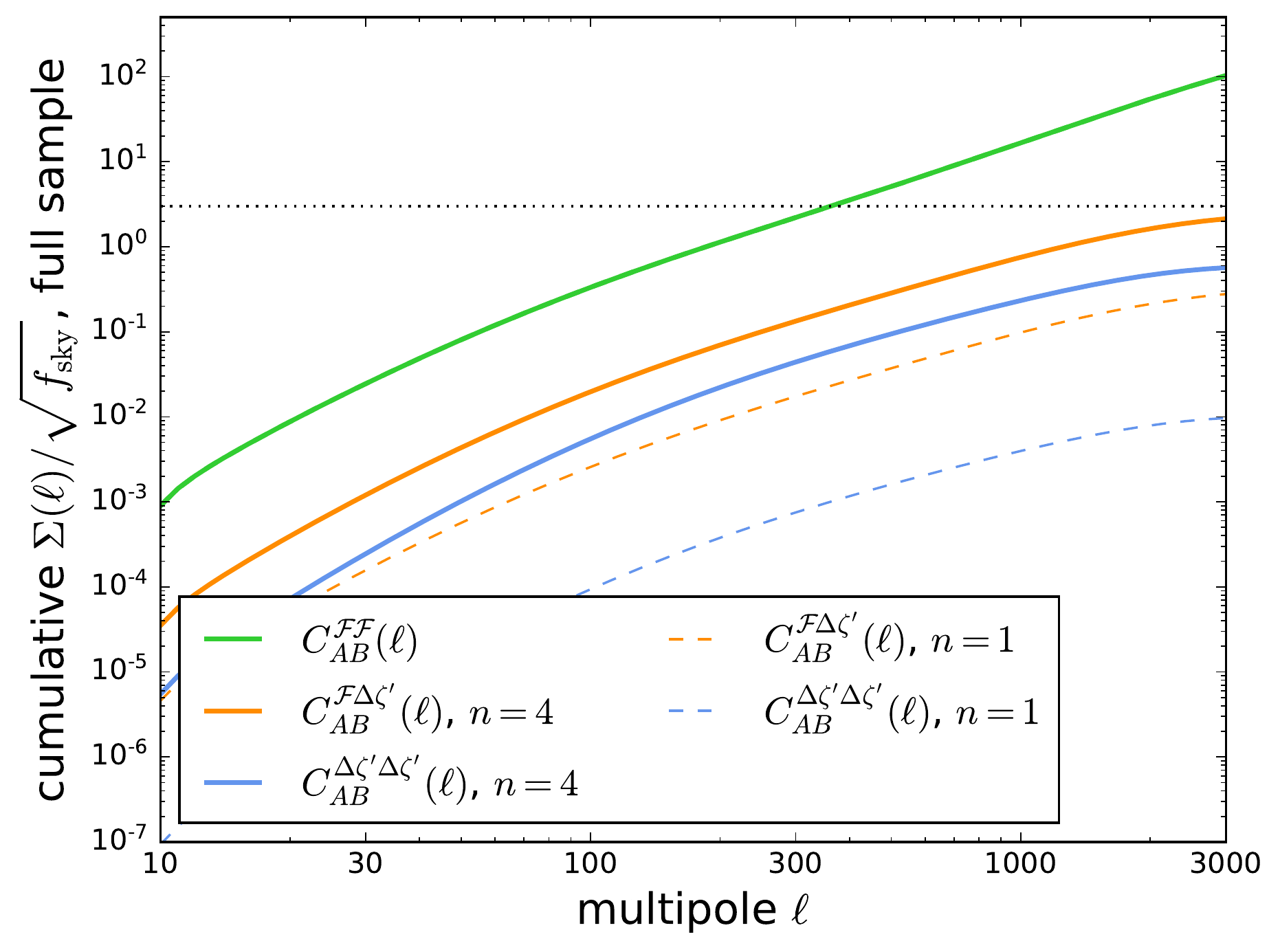}
\caption{Cumulative signal to noise-ratio $\Sigma(\ell)/\sqrt{f_\mathrm{sky}}$ for Euclid 5-bin tomography for measuring intrinsic and extrinsic flexion correlations, for two choices $n=1$ and $n=4$ for the S{\'e}rsic-index, for the full galaxy sample, i.e. without any preselection and for $\sigma_F=0.009\text{arcsec}^{-1}$.}
\label{fig:s2n_all}
\end{figure}

\begin{figure}
\centering
\includegraphics[scale=0.45]{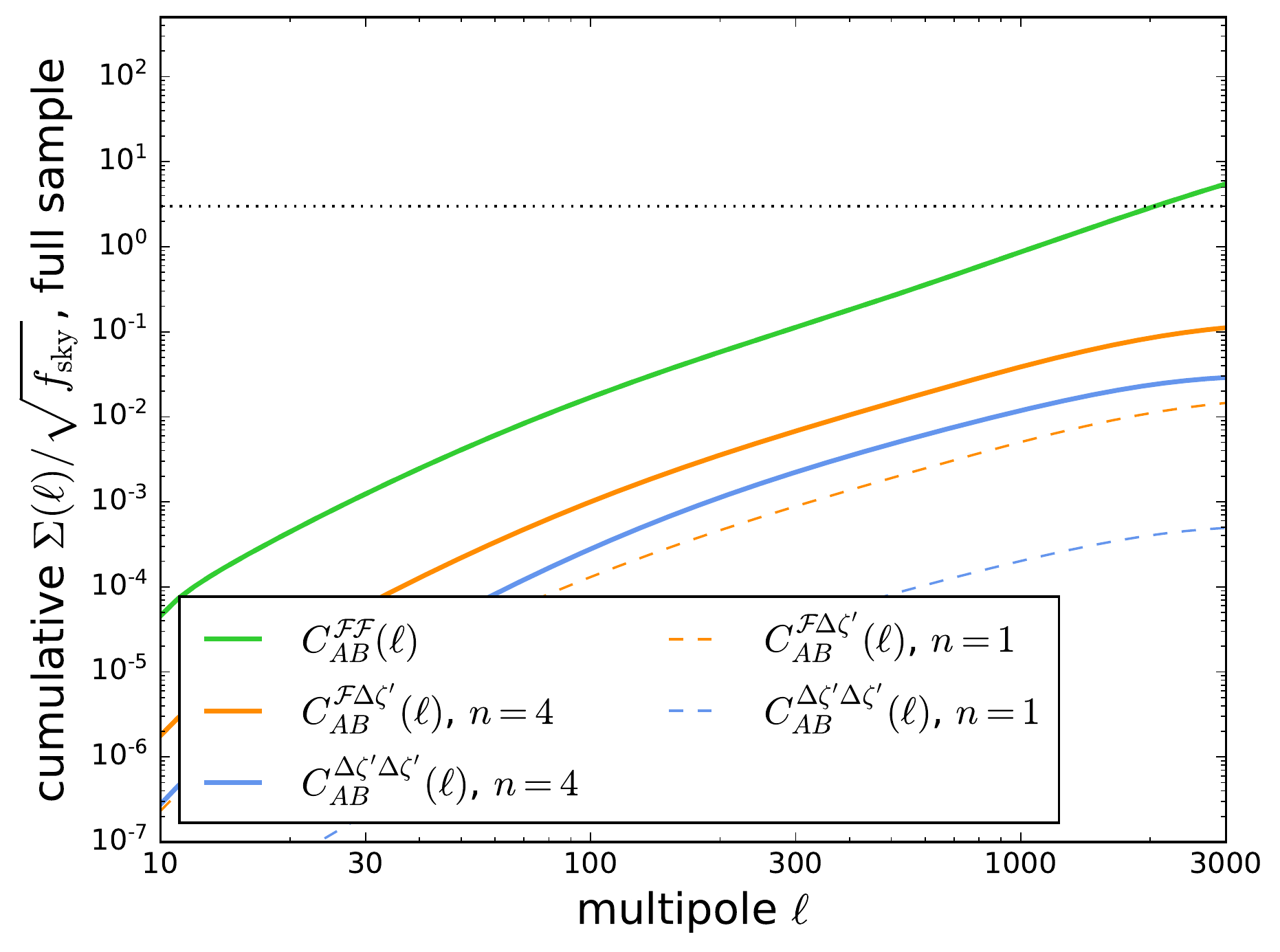}
\caption{Cumulative signal to noise-ratio $\Sigma(\ell)/\sqrt{f_\mathrm{sky}}$ for Euclid 5-bin tomography for measuring intrinsic and extrinsic flexion correlations, for two choices $n=1$ and $n=4$ for the S{\'e}rsic-index, for the full galaxy sample, and for $\sigma_F=0.04\text{arcsec}^{-1}$.}
\label{fig:s2n_all_Bacon}
\end{figure}

\begin{figure}
\centering
\includegraphics[scale=0.45]{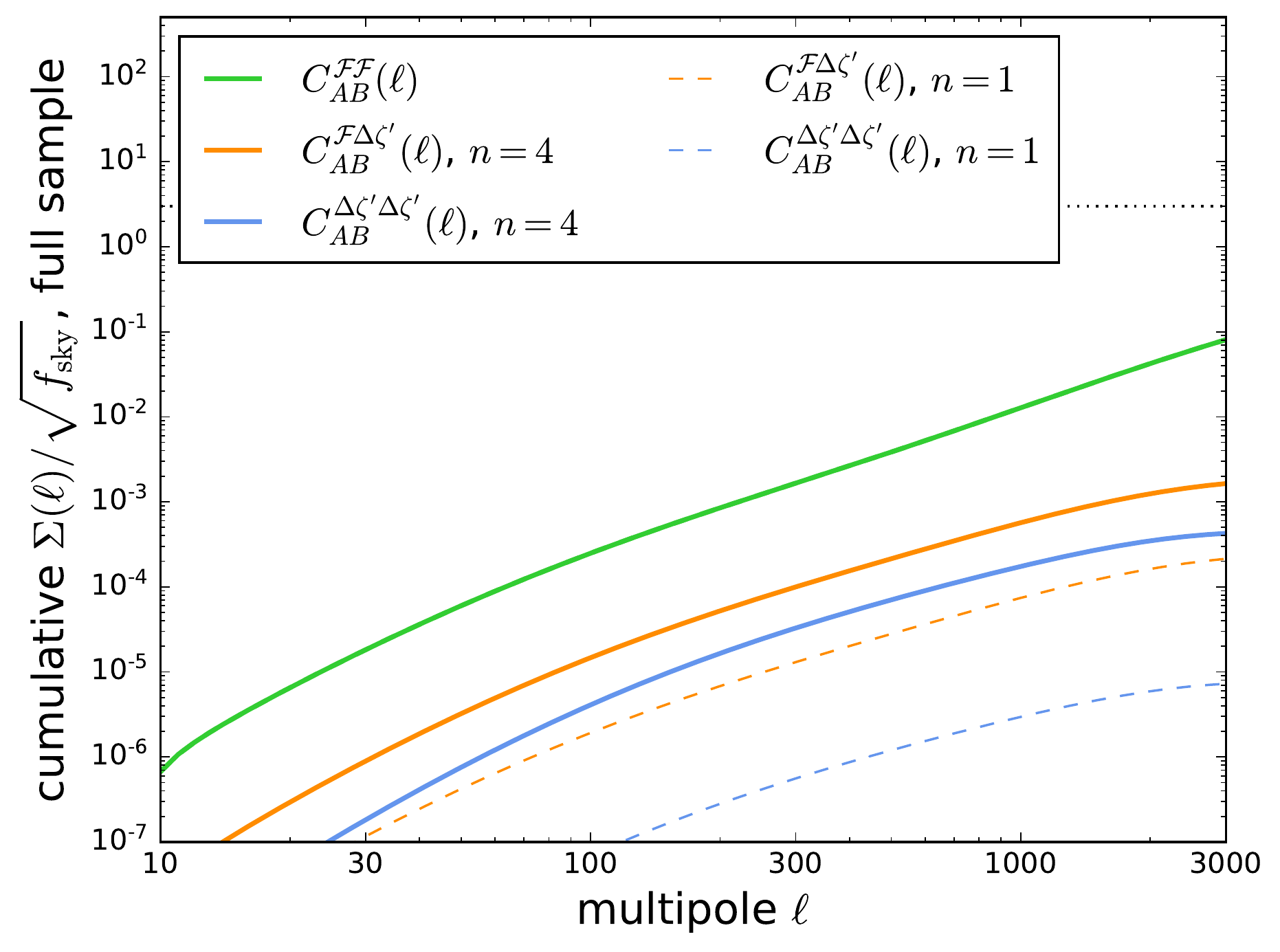}
\caption{Cumulative signal to noise-ratio $\Sigma(\ell)/\sqrt{f_\mathrm{sky}}$ for Euclid 5-bin tomography for measuring intrinsic and extrinsic flexion correlations, for two choices $n=1$ and $n=4$ for the S{\'e}rsic-index, for the full galaxy sample, and for $\sigma_F=0.33\text{arcsec}^{-1}$.}
\label{fig:s2n_all_Rowe}
\end{figure}

While in Fig.~\ref{fig:s2n_elliptical} we assumed that the elliptical galaxies were preselected, Fig.~\ref{fig:s2n_all} shows the corresponding results for the full galaxy sample of a Euclid-like survey with the dispersion of $\sigma_F=0.009 \text{arcsec}^{-1}$, also including spiral galaxies, which should not be sensitive to the proposed linear alignment model. We follow the assumption of \citet{ghosh2020intrinsic} that only one third of the sample actually consists of elliptical galaxies. Consequently, the extrinsic lensing flexions benefits directly from having a lower Poisson noise compared to the case, where ellipticals are preselected, since the sample size is thrice as large. In contrast intrinsic flexions suffer from the fact that only a fraction $q=1/3$ of the galaxies is elliptical and carries the alignment signal, leading to lower values for the signal to noise ratio for both choices of $n$, since the $GI$ spectrum is suppressed by $q$, while the $II$ contribution is even more suppressed by a factor of $q^2$. Also we present the cumulated signal to noise ratio for the dispersions of $\sigma_F=0.04 \text{arcsec}^{-1}$ and the realistic value of $\sigma_F=0.33 \text{arcsec}^{-1}$ for the whole galaxy sample, as depticted in Figs.~\ref{fig:s2n_all_Bacon} and \ref{fig:s2n_all_Rowe} respectively. Again - compared to Fig.~\ref{fig:s2n_elliptical} - these are supressed by about one order of magnitude and two to three orders of magnitude respectively. While at least the lensing flexion would be measurable for high multipole order above $\ell \approx 2000 $ for the dispersion by \citet{goldberg_galaxy-galaxy_2005}, neither the lensing nor the intrinsic flexions are measurable for the value by \citet{rowe2013} for a signal to noise ratio of $\mathrm{SNR} =100$ for the measured galaxies.

\section{Information content of intrinsic and extrinsic flexion correlations}\label{sect_fisher}
Quantifying the information content of intrinsic and extrinsic flexions with respect to cosmological parameters with the Fisher formalism, which is methodically described by \citet{tegmark_1997, amara_optimal_2007} for instance, we approximate the likelihood of the model parameters $\boldsymbol{\theta}$ given the measured data $\boldsymbol{x}(\ell)$ as a multivariate Gaussian
$\mathcal{L}\left(\boldsymbol{\theta} \vert \boldsymbol{x}(\ell)\right) \varpropto  \text{exp} \left(- \frac{1}{2}\Delta \theta^\mu F_{\mu \nu} \Delta \theta^\nu \right)$, where the Fisher matrix $F_{\mu \nu}$ is a measure of the inverse covariance of the model parameters under consideration and is derived by \citet{tegmark_1997} to be
\begin{equation}
F_{\mu \nu}  = \sum_\ell \frac{2\ell+1}{2} \text{tr} \left(\frac{\partial \mathcal{C}}{\partial \theta^\mu } \mathcal{C}^{-1} \frac{\partial \mathcal{C}}{\partial \theta^\nu } \mathcal{C}^{-1}   \right),
\end{equation}
where the derivatives of the covariance matrix with respect to the model parameters $\boldsymbol{\theta}$ are evaluated at the likelihood's best fit, and the factor $2\ell+1$ takes the multiplicity of the modes into account under the assumption of statistical isotropy. We will work with the simplifying assumption that the modes are still statistically independent even for incomplete sky coverage $f_\mathrm{sky} < 1$ but scale the Fisher-matrix with $f_\mathrm{sky}$ in the same way as the signal to noise-ratio $\Sigma$. This is a sensible simplification as most of the signal is generated on small scales.

In our analysis the covariance matrix $\mathcal{C}$ with components $\mathcal{C}_{A B}$, where $A$ and $B$ denote the redshift bin indices, is given by
\begin{equation}
\mathcal{C}_{A B}(\ell) = 
C^{\mathcal{F} \mathcal{F}}_{A B}(\ell) + C^{\Delta \zeta' \Delta \zeta'}_{A B}(\ell) + C^{\mathcal{F} \Delta \zeta'}_{A B}(\ell) + \sigma^2_{\mathcal{F}} \frac{n_{\text{tomo}}}{\bar{n}} \delta_{A B},
\end{equation}
which becomes using the results of Sect.~\ref{sect_spectra}
\begin{equation}
C_{A B}(\ell) =
\ell^6 \left[ \frac{1}{4} C^{\psi_A \psi_B} (l) + \frac{16}{9} \frac{D_{IA,\mathcal{F}}^2}{D_{IA}^2} C^{\varphi_A \varphi_B} (\ell) - \frac{2}{3}\frac{D_{IA,\mathcal{F}}}{D_{IA}} \left(C^{\psi_A \varphi_B} (\ell) + C^{\varphi_A \psi_B} (\ell)\right)   \right] + \sigma^2_{\mathcal{F}} \frac{n_{\text{tomo}}}{\bar{n}} \delta_{A B},
\end{equation}
such that it is symmetric in $A$ and $B$ as required by definition of the covariance matrix. This is ensured by symmetrisation the $GI$-component such that for $A\neq B$ either $C^{\Delta \zeta' \mathcal{F}}_{A B}(\ell) \varpropto  C^{\varphi_A \psi_B} (\ell) = 0$ for $A > B$, or $ C^{\mathcal{F} \Delta \zeta'}_{A B} (\ell)\varpropto C^{\psi_A \varphi_B}(\ell) = 0 $ for $B >A$, because lensing in lower redshift bins cannot be correlated with intrinsic alignment in higher redshift bins.

\begin{figure}
\centering
\includegraphics[scale=0.45]{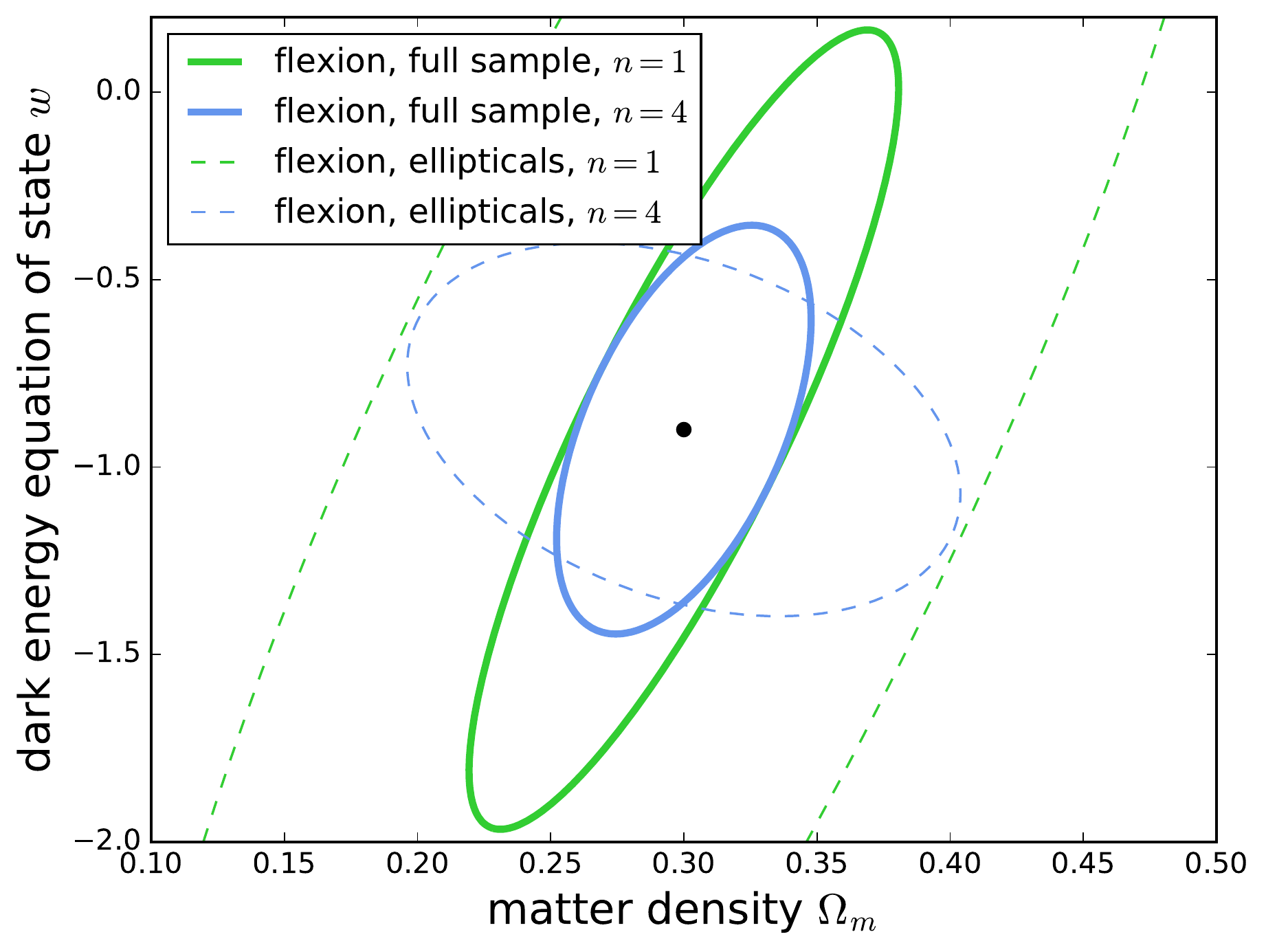}
\caption{Marginalised $1\sigma$-contours from the Fisher-matrix analysis on a standard $w$CDM-cosmology (with $w=-0.9$) for a fixed value $D=-10^{-2}$ for the alignment parameter and a smoothing scale of $10^{12}M_\odot/h$ and a flexion uncertainity of $\sigma_F = 0.009 \mathrm{arcsec}^{-1}$}: We give contours separated by S{\'e}rsic-index $n$ for both the exponential as well as the de Vaucouleurs-profile. We compare the cases for the full galaxy sample versus a sample containing only preselected elliptical galaxies.
\label{fig:fisher}
\end{figure}

\begin{figure}
\centering
\includegraphics[scale=0.45]{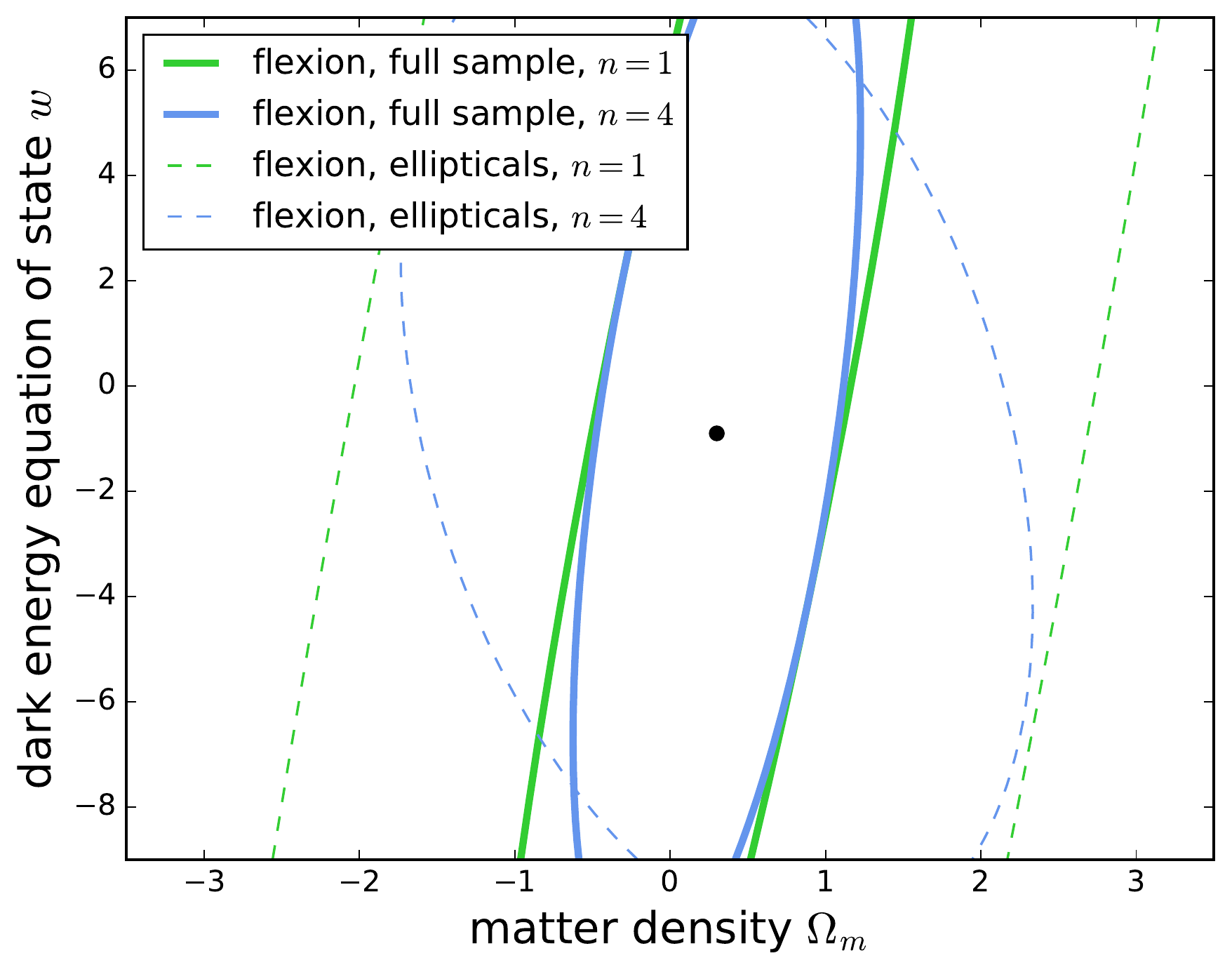}
\caption{Marginalised $1\sigma$-contours from the Fisher-matrix analysis on a standard $w$CDM-cosmology, similar to Fig.~\ref{fig:fisher}, but with $\sigma_F = 0.04 \mathrm{arcsec}^{-1}$.}
\label{fig:fisher_Bacon}
\end{figure}

Fig.~\ref{fig:fisher} then summarises the result of a Fisher-matrix analysis on the cosmological parameters $\Omega_m$ and $w$ in the context of a $w$CDM cosmology with constant equation of state parameter $w$ for the dispersion by \citet{okura_new_2007}. Those constraints are at least an order of magnitude less tight, due to the low signal strength, and therefore not competitive with the corresponding ellipticity correlations, but are in fact carrying exactly the same information about the cosmological model. Here, we also observe that the full sample selection modes have a much more constricted confidence contour since the Poissonian error is much lower while the signal strength for the $GI$-term is again weighted by $q$, while the $II$-term is surpressed by a factor of $q^2$. Furthermore, the choice of the S{\'e}rsic-index has a direct effect on the precision of the constraints: It is increased for the de Vaucouleurs-profile compared to the exponential profile, due to the increased signal strength of the intrinsic flexions. We emphasize that the point of presenting the Fisher ellipses for both Sérsic-indices, $n=1$ and $n=4$ is to present predictions for two extreme choices and to have a direct comparison between these. Most galaxies might have Sérsic indices in between these two values, such that their confidence intervals will be less constraining than for $n=4$, but more constraining than for $n=1$. However, since the dispersion by \citet{okura_new_2007} likely underestimates the uncertainites when compared to the dispersions one receives for modelations of realistic measurements by \citet{rowe2013} we also depict the cosmological constraints for the still very optimistic value of \citet{goldberg_galaxy-galaxy_2005} in Fig.~\ref{fig:fisher_Bacon} which is even an order of magnitude less constraining than the analysis for Fig.~\ref{fig:fisher}. Thus it does not have much predictive power for the parameters $\Omega_m$ and $w$, since even the lensing flexion only yields an appreciable signal on small scales. For the error of $\sigma_F = 0.33 \mathrm{arcsec}^{-1}$ by \citet{rowe2013} a prediction of reasonable constraints on the cosmological parameters is not possible, since the $1\sigma$-contours would contain a vast section of the parameter space. This is not surprising since for this error it would not even be possible to measure the lensing flexion as shown in Fig.~\ref{fig:s2n_elliptical_Rowe} and Fig.~\ref{fig:s2n_all_Rowe}. Hence the Fisher ellipses for this flexion dispersion are not depicted here.

We have verified the numerics of our code for the computation of the derivatives $\partial\ln\mathcal{C}/\partial\theta^\mu$ and the assembly of the Fisher-matrix by $(i)$ testing different finite differentiation schemes and $(ii)$ and making sure that the result converges and does not depend strongly on the step size itself. The fiducial values for the cosmological parameters correspond to the current state of knowledge on $\Lambda$CDM, where we focus on a minimal parameter set to be varied \citep[while changes in the Fisher-matrix due to varying fiducials could be tracked with the formalism outlined by][]{2016MNRAS.460.3398S}. And finally, we would like to emphasise that there is no fundamental numerical difference between inference from shear and inference from flexion spectra with two exceptions: The higher noise level and additional factors $\propto\ell^2$, both with no fundamental implications for the numerics, for both lensing and intrinsic alignments alike. We would like to emphasise that of course for very weak signals one would not expect Gaussian-likelihoods as a linearisation of the signal's dependence on the fundamental parameter is not applicable over a wide allowed parameter range: This will be the case for the flexion likelihood even for a comparatively simple model like $\Lambda$CDM. Non-Gaussian shape noise does not appear to be discussed in the literature, which is the reason why we work with a Gaussian model. The argument can be made that on small scales the signal is dominated by nonlinear structure formation with associated non-Gaussian statistics: While we take care of this by including a phenomenological extension to the CDM-spectrum, we approximate the covariance of the flexion spectra as being Gaussian.

\section{summary}\label{sect_summary}
The subject of our investigation were extrinsic and intrinsic flexions of elliptical galaxies caused by weak  lensing and intrinsic alignments. We model the effect of third derivatives of the gravitational potential in complete analogy to the effect of second derivatives of the gravitational potential which we treated in a previous study: We expose a virialised, spherically symmetric system, which is described by the Jeans-equation, to gradients in tidal gravitational fields, i.e. gravitational potentials with third derivatives. In the limit of small perturbations we can show that there is naturally a linear distortion of the stellar component which is proportional to these third derivatives, which gives rise to an intrinsic flexion-type deformation. The magnitude of these intrinsic flexions is controlled by the velocity dispersion of the galaxy and depends strongly on the stellar profile. A toy model using the family of S{\'e}rsic-profiles helps us to relate the alignment parameter for intrinsic flexions to that of intrinsic ellipticities, which we investigated in an analogous model in a previous paper.

\begin{itemize}
\item{In the linear alignment model there should be an influence of the gradients of the tidal shear field on the octupole moments of the brightness distribution, i.e. an intrinsic flexion effect. We have derived this effect in perturbation theory, relating the brightness distribution to the third derivatives of the external gravitational potential. Measuring the brightness octupole with HOLICs allows a comparison of the intrinsic flexion effect to weak lensing flexions, and to compute the angular spectra.}

\item{We derive angular spectra of the intrinsic and extrinsic (i.e. gravitational lensing) flexions including their cross-correlations by relating the flexion effect through a third angular derivative to the corresponding projected gravitational potential. By establishing an alignment potential as the projected gravitational potential weighted along the line of sight with the galaxy redshift distribution in analogy to the lensing potential, which results from the projection of the gravitational potential with the lensing efficiency function we can carry out the Limber-projection of the potential and derive the flexion spectra through prefactors of $\ell^6$. For exponential profiles, the amplitude of the spectra involving intrinsic flexions are two to three orders of magnitude below the lensing flexion spectrum.} In general, the flexion spectra are strongly increasing with multipole, reflecting the high order of the derivative of the gravitational potential that they measure. Similarly to ellipticity spectra, nonlinear structure formation increases the spectra on multipoles above $\ell\simeq300$ by one to one and a half order of magnitude.

\item{The alignment parameters which relate intrinsic ellipticity and intrinsic flexion to the second and third derivatives of the gravitational potential can be chosen consistently by considering the second and third moments of the brightness distribution, as required by the HOLIC-procedure. Assuming the S{\'e}rsic-profile family for the unperturbed brightness distribution of elliptical galaxies lets us choose the alignment parameters at second and third order in a consistent way.}

\item{The signal to noise ratios for extrinsic and intrinsic flexions in a tomographic, Euclid-like survey have been estimated with a Gaussian approximation and were found to be about $2$ for the cross-correlation between the two, and a little less for the auto-spectrum of the intrinsic flexion, when pushing the upper limit of the $\ell$-summation to $3\times 10^3$. These estimates assume a tomographic survey with five bins and a preselection of elliptical galaxies from the full sample, as well as rather large S{\'e}rsic-indices around $n=4$, corresponding to the de Vaucouleurs-profile. Also these estimates assumes a very low dispersion of $\sigma_F = 0.009 \mathrm{arcsec}^{-1}$ according to \citet{okura_new_2007}. Realistically, reachable signal to noise ratios are even lower due to higher HOLIC noise levels. Especially for the more realistic, yet still optimistic value of \citet{goldberg_galaxy-galaxy_2005} only the signal to noise ratio of about $0.1$ in the limit of $\ell=3000$ for the lensing signal is attainable, while for even higher uncertainties as $\sigma_F = 0.33 \mathrm{arcsec}^{-1}$ the signal to noise ratios would be of magnitude $10^{-3}$, and thus undetectable. The measurement of the flexion as well as its uncertainties from pixel data is an involved task, and varies strongly depending on the signal to noise ratio of the measured galaxies \citep{rowe2013}.}

\item{Computing constraints on $\Omega_m$ and $w$ in a Fisher-matrix analysis for $\sigma_F = 0.009 \mathrm{arcsec}^{-1}$ shows that there is sensitivity on these cosmological parameters contained in the intrinsic flexion signal, but that it has much lower sensitivity compared to the ellipticity signal as a consequence of the higher noise levels. For more realistic noise leves such as $\sigma_F = 0.04 \mathrm{arcsec}^{-1}$ or $\sigma_F = 0.33 \mathrm{arcsec}^{-1}$ the Fisher-matrix analysis does not yield any reliable information on the cosmological parameters contained in the intrinsic flexions. The sensitivity would be too low and the intrinsic flexions are not measurable for these noise levels. It should be emphasised that there is no fundamental physical difference between ellipticity and flexion, intrinsic or lensing-induced, in relation to their parameter sensitivity as they differ only by prefactors of $\ell$. The difference is primarily the lower signal strength and the steep functional shape of the flexion spectra, caused by the high power of $\ell$: For the full multipole range between $10$ and $3000$ considered here, the flexion signal is fully shape-noise dominated.}
\end{itemize}

Our estimates show that even in optimistic cases the intrinsic flexion could not be measured with Euclid, since shape measurements at the flexion level hardly have a precision similar to the data on the lensing cluster A1689 \citep{okura_method_2008}, even if the value for the alignment parameter is realistic and if one chooses the S{\'e}rsic index to be on the large side. For more realistic noise estimates, taking the uncertainties in measuring flexion from pixel data into account \citep{rowe2013} one can not even measure the lensing flexion for optimistic noise levels. It is unclear by what mechanism there could be an effect on the third moments of the brightness distribution of spiral galaxies in analogy to the linear alignment model for elliptical galaxies discussed here. From this point of view one would need to resort to the case that all galaxies potentially carry the lensing flexion signal, but only a subset of them the intrinsic flexion signal.

\section*{Acknowledgements}
ESG would like to thank Studienstiftung des Deutschen Volkes for financial support. BG thanks Indian Institute of Science for providing her with the CV Raman Postdoctoral fellowship. The authors would like to express their gratitude to David Bacon for his interest and his critical questions, and are also very grateful to Ruth Durrer for her time to discuss the contents of the paper, as well as her helpful comments and critical remarks. The authors kindly thank the anonymous referee for their time to review this work as well as many suggestions for improvement.

\section*{Data Availability}
There are no new data associated with this article.

\bibliographystyle{mnras}
\bibliography{references}

\appendix

\section{Derivation of the flexion decomposition in Gravitational Lensing}\label{appA}
In this section it is shown how one can derive the basis decomposition of the third derivative of the Newtonian potential $\Phi_{abc}$ in terms of $4\times4$ matrices stated in eqn. (\ref{eq:Diracs}) from the flexion decompostion of gravitational lensing. In fact the rank-3 object $\Phi_{abc}$ should have the same decomposition as the rank-3-object $D_{ijk} \equiv \partial_{\theta_k}  \mathcal{A}_{ij} =-\psi_{kij} $ quantifying the flexion of weak gravitational lensing as the third derivative of the lensing potential $\psi$ (for reference confirm \citet{bacon_weak_2006, goldberg_galaxy-galaxy_2005}). To find the decompostion we in principle follow the derivation of the lensing flexion in \citet{bacon_weak_2006} but express everything in terms of the decompostion into Pauli matrices of the Jacobian $\mathcal{A}=\partial \boldsymbol{\beta} / \partial \boldsymbol{\theta}$ of the lens map (\citet{Bartelmann2001,Bartelmann2017WeakGL})
\begin{equation}
\mathcal{A}= \begin{pmatrix}1- \kappa -\gamma_{+} & -\gamma_{\times} \\ -\gamma_{\times}  & 1- \kappa +\gamma_{+} \end{pmatrix} =\mathbb{1}_2 -\kappa \sigma^{(0)} - \gamma_{+} \sigma^{(3)} - \gamma_{\times} \sigma^{(1)} .
\end{equation}
thereby separating the flexion contributions into linear independent terms. Here $\kappa = 1/2\left(\psi_{00}+\psi_{11} \right)$ denotes the convergence, and $\gamma_{+} =1/2\left(\psi_{00}-\psi_{11} \right)$ and $\gamma_{\times} =\psi_{01} = \psi_{10}$ the shear components. The rank-3 tensor $D_{ijk}$ considered here is symmetric in all its three indices due to the interchangeability of the partial derivatives, thus it has four degrees of freedom. As discussed in \citet{schafer_validity_2012} it can be represented by a $4\times4$ block diagonal matrix $\left(\mathcal{D}\right)_{i+2k, j+2k} = D_{ijk}$ with
\begin{equation}
\begin{split}
\mathcal{D}=& \begin{pmatrix}  D_{ij0} & 0 \\0 &  D_{ij1} \end{pmatrix} =  \begin{pmatrix}   \partial_0  \mathcal{A}_{ij} & 0 \\0 &   \partial_1  \mathcal{A}_{ij} \end{pmatrix}
= -  \begin{pmatrix} \partial_0 \kappa \sigma^{(0)} +  \partial_0 \gamma_{+} \sigma^{(3)} +  \partial_0 \gamma_{\times} \sigma^{(1)} & 0 \\0 &   \partial_1 \kappa \sigma^{(0)} +  \partial_1 \gamma_{+} \sigma^{(3)} +  \partial_1 \gamma_{\times} \sigma^{(1)} \end{pmatrix} \,.
\end{split}
\end{equation}
Since $\partial_0 \kappa = \partial_0 \gamma_{+} +\partial_1  \gamma_{\times}$ and $\partial_1 \kappa =  -\partial_1 \gamma_{+} + \partial_0 \gamma_{\times}$
hold, the required four degrees of freedom of $\mathcal{D}$ can solely be expressed by derivatives of the shear components $ \partial_0 \gamma_{+}$, $\partial_1 \gamma_{+}$, $ \partial_0 \gamma_{\times}$ and $ \partial_1 \gamma_{\times}$. Consequently, it follows for $\mathcal{D}$, that
\begin{equation}
\begin{split}
\mathcal{D} = -\partial_0 \gamma_{+} \begin{pmatrix} \sigma^{(0)} + \sigma^{(3)}  & 0 \\0 &   0\end{pmatrix} -  \partial_1 \gamma_{+} \begin{pmatrix}0  & 0 \\0 &  \sigma^{(3)} - \sigma^{(0)} \end{pmatrix} -\partial_0 \gamma_{\times} \begin{pmatrix} \sigma^{(1)}  & 0 \\0 & \sigma^{(0)} \end{pmatrix}  -\partial_1 \gamma_{\times} \begin{pmatrix} \sigma^{(0)}  & 0 \\0 & \sigma^{(1)} \end{pmatrix}
\end{split}
\end{equation}
Using the definitions of the two complex flexion fields
\begin{equation}
\mathcal{F} =\mathcal{F}_1 + \mathrm{i}\mathcal{F}_2  = \frac{1}{2} \partial \partial^* \partial \psi = \partial \kappa = \partial^* \gamma\, \quad \text{and} \quad \mathcal{G} = \mathcal{G}_1 + \mathrm{i}\mathcal{G}_2 =\frac{1}{2} \partial \partial \partial \psi= \partial \gamma\,,
\end{equation}
introduced in \citet{bacon_weak_2006} with $\partial = \partial_0 + \mathrm{i} \partial_1 $ being the complex derivative and its complex conjugate $\partial^* = \partial_0 - \mathrm{i} \partial_1 $ one can express $\mathcal{D}$ in terms of these four different degrees of freedom which are distinguished by different spins: Under rotation with angle $\phi$ they transform as $\mathcal{F} \rightarrow \mathcal{F} \exp\left( \mathrm{i} \phi\right)$ and $\mathcal{G} \rightarrow \mathcal{G} \exp\left( \mathrm{i}3 \phi\right)$, as discussed in \citet{bacon_weak_2006}.
Thus the derivatives of the shears can be expressed as
\begin{align}
\partial_0 \gamma_{+} = \frac{\mathcal{F}_1 + \mathcal{G}_1}{2}\,, \partial_1 \gamma_{+} = \frac{\mathcal{G}_2 - \mathcal{F}_2}{2} \quad \text{and} \quad \partial_0 \gamma_{\times} = \frac{\mathcal{F}_2 + \mathcal{G}_2}{2}\,,\partial_1 \gamma_{\times} = \frac{\mathcal{F}_1- \mathcal{G}_1}{2}\,.
\end{align}
This leads to the following possible decomposition for the result of \citet{bacon_weak_2006}:
\begin{equation}\label{eq:decomp1}
\begin{split}
\mathcal{D} =& -\frac{1}{2} \mathcal{F}_1 \begin{pmatrix} 2\sigma^{(0)} + \sigma^{(3)}  & 0 \\0 & \sigma^{(1)}\end{pmatrix} -\frac{1}{2} \mathcal{F}_2 \begin{pmatrix} \sigma^{(1)}  & 0 \\0 &  2\sigma^{(0)} - \sigma^{(3)} \end{pmatrix} -\frac{1}{2} \mathcal{G}_1 \begin{pmatrix}  \sigma^{(3)}  & 0 \\0 & -\sigma^{(1)}\end{pmatrix} -\frac{1}{2} \mathcal{G}_2 \begin{pmatrix} \sigma^{(1)}  & 0 \\0 & \sigma^{(3)} \end{pmatrix}\,.
\end{split}
\end{equation}
One can explicitly confirm that these matrices form an orthogonal basis by contraction,
using that the Pauli matrices themselves form an orthonormal basis and are traceless, i.e.
$\sigma^{(i)} \sigma^{(j)}= \mathbb{1}_2 \delta_{i j} + \mathrm{i} \epsilon_{ijk}\sigma^{(k)}$ and $\text{Tr}{\left(\sigma^{(k)}\right)} =0$ hold.
In a next step one should perform a normalisation of the basis matrices $\Delta^{(n)}$ such that the distinct degrees of freedom of the flexion field, denoted as $d_n$, can be extracted from the following projection
\begin{equation}
d_n= \frac{1}{4} \text{Tr}\left(\mathcal{D} \cdot \Delta^{(n)} \right)\,, \quad \text{with $n=1,2,3,4$.}
\end{equation}
Since for the matrices of the spin-1 field one receives
\begin{equation}
\begin{split}
\frac{1}{4} \text{Tr}\left(\begin{pmatrix} 2\sigma^{(0)} + \sigma^{(3)}  & 0 \\0 & \sigma^{(1)}\end{pmatrix} \cdot \begin{pmatrix} 2\sigma^{(0)} + \sigma^{(3)}  & 0 \\0 & \sigma^{(1)}\end{pmatrix}  \right) = 3 \quad \text{and} \, \frac{1}{4} \text{Tr}\left(\begin{pmatrix} \sigma^{(1)}  & 0 \\0 & 2\sigma^{(0)} - \sigma^{(3)}\end{pmatrix} \cdot \begin{pmatrix}\sigma^{(1)}  & 0 \\0 & 2\sigma^{(0)} - \sigma^{(3)}\end{pmatrix}  \right) = 3
\end{split}
\end{equation}
we introduce a normalisation factor $1/\sqrt{3}$ in front of the basis matrices of the $\mathcal{F}$-flexion field. Then the orthonormal decomposition of $D_{ijk}$ is given by
\begin{equation}
\begin{split}
\mathcal{D} = d_1 \frac{1}{\sqrt{3}} \begin{pmatrix} 2\sigma^{(0)} + \sigma^{(3)}  & 0 \\0 & \sigma^{(1)}\end{pmatrix}+d_2  \frac{1}{\sqrt{3}} \begin{pmatrix} \sigma^{(1)}  & 0 \\0 &  2\sigma^{(0)}- \sigma^{(3)} \end{pmatrix}+d_3\begin{pmatrix}  \sigma^{(3)}  & 0 \\0 & -\sigma^{(1)}\end{pmatrix}+d_4\begin{pmatrix} \sigma^{(1)}  & 0 \\0 & \sigma^{(3)} \end{pmatrix}\,,
\end{split}
\end{equation}
with coefficients $d_1 \equiv  -\sqrt{3}/2 \, \mathcal{F}_1$, $d_2 \equiv  -\sqrt{3}/2\, \mathcal{F}_2$, $d_3\equiv  -1/2\, \mathcal{G}_1$ and $d_4\equiv  -1/2 \,\mathcal{G}_2$. To conclude $\mathcal{D}$ can thus be decomposed using four different matrices $\Delta^{(n)}$, where two of them are required for the spin-1 field $\mathcal{F}$ and the other two for the spin-3 field $\mathcal{G}$:
\begin{align}
\text{spin-1-types:} \,\, \Delta^{(1)} = \frac{1}{\sqrt{3}} \begin{pmatrix} 2 \sigma^{(0)} + \sigma^{(3)} & 0 \\ 0 & \sigma^{(1)}    \end{pmatrix}, \, \Delta^{(2)} = \frac{1}{\sqrt{3}} \begin{pmatrix} \sigma^{(1)} & 0 \\ 0 & 2 \sigma^{(0)} - \sigma^{(3)}\end{pmatrix},
\quad \,
\text{spin-3-types:} \,\, \Delta^{(3)} = \begin{pmatrix} \sigma^{(3)} & 0 \\ 0 & -\sigma^{(1)}    \end{pmatrix}, \,  \Delta^{(4)}= \begin{pmatrix} \sigma^{(1)} & 0 \\ 0 & \sigma^{(3)}    \end{pmatrix}.
\end{align}
Now, this decomposition for the lensing flexions can also be used for the intrinsic flexions, since both of them are basically derived from the third derivative of the gravitational potential $\Phi_{ijk}$.


\bsp
\label{lastpage}
\end{document}